\documentclass[aps, physrev, amsmath, amssymb, superscriptaddress]{revtex4-2}

\makeatletter
\def\switch@array{}
\makeatother

\usepackage[T1]{fontenc}
\usepackage{array}
\usepackage{tabularx}
\newcolumntype{Y}{>{\raggedright\arraybackslash}X}
\usepackage{booktabs}
\usepackage{amsmath}

\usepackage{graphicx}
\usepackage{multirow}
\usepackage{bm}
\usepackage{hyperref}
\usepackage{float}
\usepackage{tikz}
\usepackage{comment}
\usepackage{todonotes}
\usetikzlibrary{calc, angles, quotes}

\begin{document}

\title{\textbf{Feasibility of satellite-augmented global quantum repeater networks} 
}%

\author{Manik Dawar}
\affiliation{Airbus Central Research and Technology, 82024 Ottobrunn, Germany
}%
\affiliation{
 Institute for Quantum Physics and Center for Optical Quantum Technologies, University of Hamburg,
Luruper Chaussee 149, 22761 Hamburg, Germany
}%

\author{Clement Paillet}
\affiliation{%
 Airbus Defence and Space, Toulouse, France
}%

\author{Nilesh Vyas}
\affiliation{%
 Airbus Central Research and Technology, 82024 Ottobrunn, Germany
}%

\author{Andrew Thain}
\affiliation{%
 Airbus Defence and Space, Toulouse, France
}%

\author{Rodrigo Henriques Guilherme}
\affiliation{%
 Airbus Defence and Space, Toulouse, France
}%

\author{Ralf Riedinger}
\affiliation{
 Institute for Quantum Physics and Center for Optical Quantum Technologies, University of Hamburg, Luruper Chaussee 149, 22761 Hamburg, Germany
}%

\begin{abstract}
A large scale quantum network requires the distribution of high-fidelity end-to-end entanglement. To overcome the range limitations inherent to terrestrial fiber, a leading architecture has emerged: satellite-based sources transmitting entanglement to quantum repeaters on the ground. By bridging the gap between abstract analytical frameworks and computationally heavy numerical simulations, this paper provides the first quantitative answer to the question of such a network's achievable performance with current and near-term space technology, while accounting for entanglement swapping and purification. This is achieved by integrating a detailed physical model of a satellite-to-ground link into an analytical entanglement resource estimation framework for quantum repeaters, enabling an optimization of the end-to-end entanglement rate.
Our analysis, performed across leading quantum hardware platforms, shows that Low Earth Orbit satellite constellations combined with quantum repeaters employing Neutral Atom or Nitrogen and Silicon Vacancy qubits, could enable a global quantum network, distributing entanglement over distances up to 20,000 km, sufficient for connecting any two points on Earth. This work highlights the major bottlenecks in space and quantum hardware technologies, which need to be addressed, thereby guiding informed investments necessary for enabling a large scale quantum network. 
\end{abstract}
\maketitle

\section{Introduction}
\label{sec:one}
The vision of a large scale quantum network \cite{wehner2018quantum, caleffi2021quantum}, capable of connecting quantum processors for enabling secure communication, distributed sensing, and distributed quantum computing, hinges on our ability to distribute high-fidelity entanglement over large distances \cite{briegel1998quantum}. Research has historically advanced along two primary avenues: terrestrial fiber-based networks and free-space satellite links. 

While terrestrial networks leveraging quantum repeaters provide a path for overcoming the exponential photon loss in optical fibers \cite{bruzewicz2019trapped, wintersperger2023neutral, pezzagna2021quantum, nguyen2019quantum}, scaling them to global distances remains a formidable resource challenge \cite{dawar2024quantum}. Concurrently, satellite-based quantum communication has emerged as a promising architecture. By transmitting photons through the vacuum of space, satellite links bypass thousands of kilometers of lossy fiber \cite{PLOB2017, takeoka2014fundamental}. Landmark experiments with the Micius satellite have successfully demonstrated the viability of this approach \cite{Micius_Entanglement}. It has, however, its own limitations. The performance of satellite-to-ground links can vary drastically due to weather, lighting conditions, and significant signal losses at small elevation angles. Line of sight availability is also limited by orbital mechanics, restricting link duration. A reliable quantum network must therefore utilise the strengths of both, the terrestrial and satellite approaches.

We consider a hybrid quantum network architecture consisting of satellites transmitting entangled photons to quantum repeaters on the ground (see Figure~\ref{fig:sat_repeater}). These ground stations use entanglement purification and swapping protocols in a nested fashion to provide an end-to-end entanglement service between requested nodes \cite{azuma2023quantum, muralidharan2016optimal}. This paper presents, to our knowledge, the first work to provide concrete performance estimates for such a satellite-augmented quantum repeater network while explicitly accounting for entanglement purification. A second novelty is a direct comparative analysis between the most promising quantum hardware technologies for large-scale networks: Nitrogen Vacancy (NV) \cite{gali2009theory}, Silicon Vacancy (SiV) color centers in diamond, and Neutral Atom qubits \cite{saffman2019quantum}.
  
These novelties are enabled by extending the work of Dawar et.~al.~\cite{dawar2024quantum}, by replacing their abstract elementary link with a detailed, multi-parameter physical model of a satellite-based entanglement distribution link. We leverage this integrated model to perform a joint optimization of three key operational parameters: the mean photon-pair number ($\mu$) of the entangled photon source on the satellites, the distance ($L_0$) between adjacent Optical Ground Stations (OGS), and the target fidelity ($F_t$) that the network must maintain. A simultaneous optimization of these interdependent parameters enables a quantitative estimation of a global quantum network's performance. We repeat the analysis for three scenarios representing the expected levels of maturity of the space technologies used in such a network: state of the art technology (Scenario A), near-term technology realizable within 5-10 years (Scenario B), and futuristic technology, realizable in 10-15 years (Scenario C). 

The remainder of this paper is structured as follows. In Section \ref{sec:related work}, we briefly summarise related work to better contextualise the relevance of this paper. In Section \ref{sec:network model}, we elucidate our basic assumptions about the network architecture. In Section \ref{sec:Joint optimization: maximising network performance}, we introduce the optimization problem of maximizing the network performance. In Section \ref{sec:results}, we present our performance estimates for such an optimized large scale quantum network across the three aforementioned scenarios. Finally, in Section \ref{sec:discussion and conclusion}, we present the insights gained from the analysis in this paper and conclude.

\section{Related Work}
\label{sec:related work}
The ambitious pursuit of a large scale quantum network \cite{wehner2018quantum, caleffi2021quantum} has ignited foundational research across physics, engineering, and computer science. However, the path from theoretical vision to practical reality is marked by significant challenges in modeling accuracy, integration, and scalability. The community's approach has largely been bifurcated, creating a dichotomy between high-level analytical frameworks \cite{pirandola2019end, pirandola2017fundamental_limits} and detailed, low-level numerical simulations \cite{coopmans2021netsquid, wu2021sequence, wallnofer2022simulating}.

Analytical models establishing end-to-end capacity bounds must abstract away hardware-specific variability, creating a gap between theoretical potential and achievable performance \cite{pirandola2019end, takeoka2014fundamental, PLOB2017}. This has motivated deep dives into optimizing specific repeater architectures \cite{muralidharan2016optimal, mantri2025comparing, GuhaRateLossQR, He21} and protocols for entanglement swapping and purification \cite{shchukin2022optimal, jones2016design, galetsky2025feasibility}. 

This complexity has spurred quantum network simulation platforms like NetSquid \cite{coopmans2021netsquid}, SeQUeNCe \cite{wu2021sequence}, QuNetSim \cite{diadamo2021qunetsim}, and SimulaQron \cite{dahlberg2018simulaqron}, to name a few. These platforms could be applied for detailed hardware analysis but are computationally impractical for exploring the vast parameter space of a global network. This has led to a new generation of simulations and analysis focused on the unique challenges of satellite-based quantum communication \cite{gundougan2021proposal, liorni2021repeaters, pirandola2021satellite}. Researchers now model dynamic, multi-satellite constellations \cite{wallnofer2022simulating, meister2025simulation} and connect metropolitan \textit{Quantum Cities} \cite{yehia2024connecting} to benchmark performance with near-term technology. These high-fidelity simulations \cite{wallnofer2024faithfully} are crucial but are fundamentally tools for analysis, not for discovering globally optimal designs. 

This landscape reveals a critical, yet largely unaddressed gap. While simulators offer high-fidelity models for specific architectures, they cannot efficiently perform a system-wide optimization over the numerous interdependent parameters that characterize a network. Conversely, frameworks for the analysis of generic repeater chains \cite{kamin2023exact, goodenough2021optimizing, haldar2024fast} are scalable but lack the physical detail to capture the unique characteristics of a satellite-to-ground link \cite{hirche2023quantum}. Ji et.~al.~\cite{ji2025global} combine a detailed model for space to ground links from LEO satellites, along with their orbital dynamics, with single atom memories in optical cavities on the ground. The quantum information from the photons is mapped onto the atomic memories on the ground using Cavity Assisted Photon Scattering (CAPS), and deterministic entanglement swapping is performed using Rydberg gates. The model, however, does not consider entanglement purification or any error correction for counteracting the degradation of fidelities due to decoherence and imperfect entanglement swapping.  

Our work directly bridges these gaps. We advance the analysis of \cite{dawar2024quantum} by integrating a detailed, multi-parameter physical model of a satellite-based entanglement link directly into their resource estimation framework for quantum repeater chains performing both, entanglement swapping and purification. This unification enables a joint multi-parameter optimization of the network performance across the entire system. Rather than being limited to optimizing subsystems in isolation, we can now simultaneously tune the source's mean photon-pair number $\mu$ per pulse, the inter-station hop length ($L_0$), and the repeater protocol's target fidelity ($F_t$) to achieve the maximum possible end-to-end entanglement rates at any given distance. 

\section{Network model}
\label{sec:network model}
\begin{figure}[htbp]
    \centering
    \includegraphics[width=0.9\textwidth]{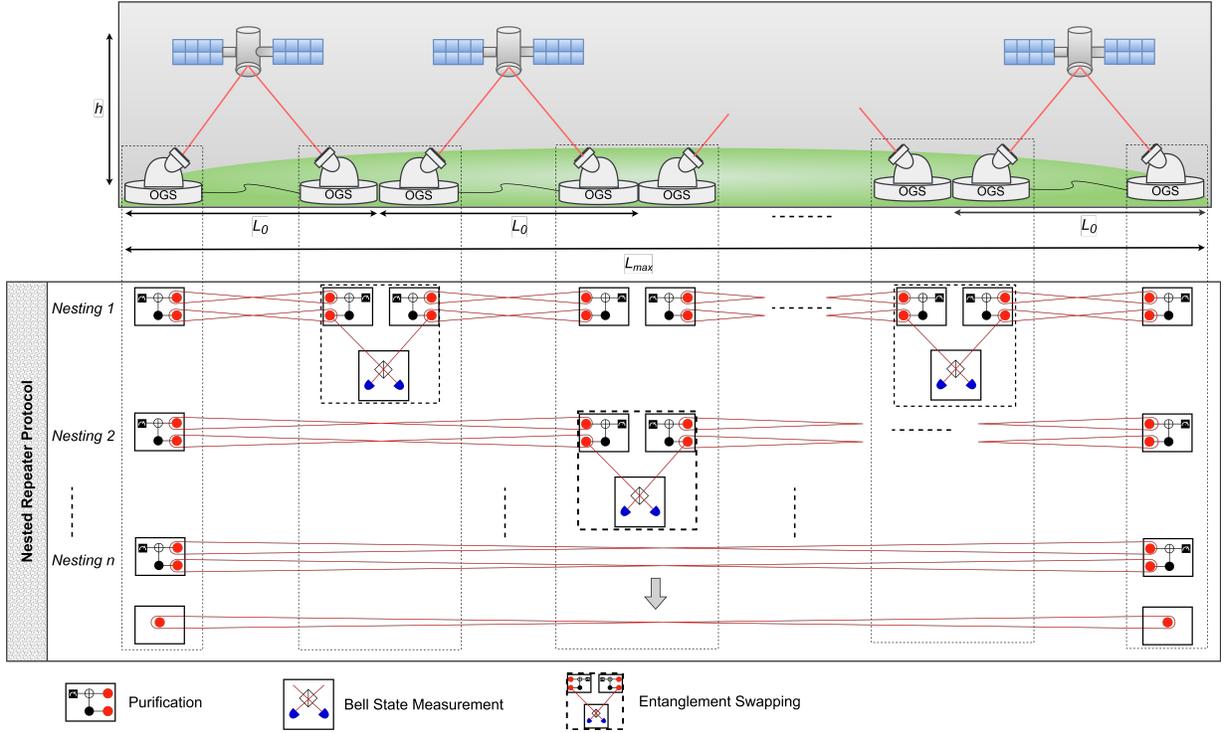}
    \caption{Satellite-assisted nested repeater protocol. SPDC sources mounted on LEO satellites send entangled photons to the optical ground stations (OGS). Once the entanglement is stored in the quantum memories at the OGS, the nested repeater protocol (Appendix \ref{app:Theoretical Framework for Repeater-Assisted Networks}) is employed to establish a direct entanglement link between the end nodes.}
    \label{fig:sat_repeater}
\end{figure}

We consider a constellation of satellites orbiting over a collection of Optical Ground Stations (OGSs) (see Figure~\ref{fig:sat_repeater}). 
An entangled photon pair is generated on the satellite by a Spontaneous Parametric Down-Conversion (SPDC) source operating at an average of $\mu$ photons per pulse. From its creation, the entanglement's fidelity is imperfect due to potential for multi-pair emissions \cite{takeoka2015full, Ma2012, caminati2006nonseparable}, which act as an intrinsic noise source. Specifically, the fidelity $F_\text{init}$ goes down with as
\begin{equation}
    F_\text{init} = \frac{2-3\text{QBER}(\mu)}{2}
    \label{eq:f init main}
\end{equation}
where $\text{QBER}(\mu)$ is the Quantum Bit Error Rate. \eqref{eq:f init main} follows directly from the definition of a Werner state (which we use to model the entangled pair arriving at the OGSs) with a QBER
\begin{equation}
    \text{QBER} = e_0 - \frac{(e_0-e_d) \eta_A \eta_B \mu (1+\frac{\mu}{2})}
    {Q_\mu (1+\eta_A\frac{\mu}{2})(1+\eta_B\frac{\mu}{2})(1+\eta_A  \frac{\mu}{2}+\eta_B\frac{\mu}{2}-\eta_A \eta_B\frac{\mu}{2})}
    \label{eq:qber main}
\end{equation}
$e_0$ is the error rate of background counts, $e_d$ is the probability for a Bell state to hit the wrong detector. $\eta_A$ and $\eta_B$ are the satellite-ground transmission efficiencies for each of the two entangled photons. In the rest of the paper, we assume a symmetric channel: $\eta_A = \eta_B =: \eta$, for simplicity. For a derivation of \eqref{eq:qber main}, see Appendix \ref{app:error models}.

The photons are transmitted from the satellite towards two separate OGS, facing several sources of loss: diffraction causes the beam to spread, pointing errors result from imperfect satellite tracking $(\eta_\text{diff-BPE})$, and atmospheric loss occurs from absorption and scattering $(\eta_\text{atm})$. 
Upon arrival, the signal suffers from coupling loss $(\eta_\text{cpl})$ when entering optical fibers via the ground telescope and corruption from stray background light. See Appendix \ref{app:error models} for a description of these errors. 

The resultant transmission efficiency $\eta$ of the satellite to ground channel is therefore
\begin{equation}
    \eta(L_0) = \eta_\text{diff-BPE}(L_0) \eta_\text{atm}(L_0) \eta_\text{cpl}
    \label{eq: transmission efficiency main}
\end{equation}
where $L_0$ is the distance between adjacent ground stations. We define these component efficiencies in detail in Appendix \ref{app:error models}. Future works could incorporate dynamic satellite orbits and extend to asymmetric space-to-ground channels with $\eta_A \neq \eta_B$. 

The entangled photons are absorbed into the quantum memories on the ground using the CAPS protocol \cite{kikura2025passive}, which succeeds with probability $\eta_{\text{CAPS}}$. Once CAPS succeeds, the entanglement links can be lengthened using entanglement swapping or enhanced using entanglement purification. We consider deterministic gate-based implementations of swapping and purification, which consequently inherit gate and measurement errors from the quantum computing platforms used on the ground nodes \cite{dawar2024quantum}. We elaborate our treatment of entanglement swapping and purification with gate and measurement errors in Appendix \ref{app:Theoretical Framework for Repeater-Assisted Networks}.

We consider a nested repeater protocol, like the one shown in Figure~\ref{fig:sat_repeater}: pairs of adjacent links are joined using entanglement swapping in each nesting level and purified. Because of this, the length of an elementary entanglement link doubles with each nesting level. Therefore, for $D+1$ nodes, $\log_2D$ nesting levels are required to establish an entanglement between the end nodes. This consumes $D^\lambda$ elementary entanglements, where $\lambda$ is the resource scaling exponent \cite{briegel1998quantum}.

The aim is to have an end-to-end entanglement with fidelity $F_t$. However, entanglements with fidelity $F_\text{init}$ in general $\neq F_t$ are received on the ground. If $F_\text{init} < F_t$, then the entanglements are purified -- $m$ times -- until their fidelity becomes greater or equal to $\geq F_t$. If, however, $F_\text{init} > F_t$, then pairs of adjacent links are swapped -- along $s$ levels -- until the fidelity of the swapped entanglements falls below $F_t$, which are then again purified to $\geq F_t$. After applying either of the above two conditions or if $F_\text{init} = F_t$ already, the nested protocol is applied. 

Dawar et.~al.~\cite{dawar2024quantum} argued that a finite coherence time $T_2$ of a quantum memory limits the number of nodes $D^* + 1$ in a repeater chain along which entanglement could be distributed at a desired end-to-end fidelity $F_t$, using such a nested protocol -- see Equation (21) in \cite{dawar2024quantum}:
\begin{equation}
    D^*(\mu, L_0, F_t) \approx 2^{s(\mu, L_0, F_t)} \left( \left( \frac{2}{\tilde{P}_F(\mu, L_0, F_t)}\right)^{-m(\mu, L_0, F_t)}R_0(\mu, L_0) T_2 \sqrt{-\log{\left(\frac{3 \sqrt{\frac{4 F_{l} - 1}{4 \eta_{\text{ro}}^{2} - 1}} + 1}{4 F_t} \right)}}\right)^{\frac{1}{\lambda(F_t)-1}} 
    \label{eq:final_dstar}
\end{equation}
$\Tilde{P}_F$ is the multiplicative mean of the purification acceptance probabilities, such that $\Tilde{P}_F^m \approx \prod_1^m P_{F_i}$, where $P_{F_i}$ is the purification success probability at the $i$th purification level -- see Equation (15) in \cite{dawar2024quantum}, $F_l$ is a fidelity fixed-point of purification, below which, purification reduces the fidelity instead of improving it (see Equation \eqref{eq:purification eta eg 1} for the purification map) and $\eta_{\text{ro}}$ is the read-out measurement error of the qubits stored in the quantum memory. Finally, $R_0$ is the generation rate of elementary entanglements among adjacent nodes:
\begin{equation}
    R_0 = N_m f \beta \eta_{\text{CAPS}} \eta_{\text{demux}}^2\eta_{\text{detect}} Q_\mu (\mu, L_0)
    \label{eq:ultimate r0}
\end{equation}
$N_m$ is the number of multiplexing modes, $f$ is the emission rate of the entangled photon source and  $\mu$ is the expectation value of the number of photons per temporal mode. We note that the source bandwidth $f$ is limited by the spin-photon interface bandwidth $f_s$. $\beta$ is consequently the ratio between the interface and source bandwidths $f_s/f$. $\eta_{\text{CAPS}}$ is the efficiency of the CAPS protocol, $\eta_{\text{demux}}$ is the demultiplexing efficiency, $\eta_{\text{detect}}$ is the detection efficiency for the CAPS protocol and $Q_\mu$ is the source gain (see Appendix \ref{app:mean photon number}), which takes into account, the overall channel transmission efficiency $\eta$ \eqref{eq: transmission efficiency main}.

The total achievable network distance $L^*$ is thence:
\begin{equation}
    L^*(\mu, L_0, F_t) \approx D^*(\mu, L_0, F_t) \times L_0
    \label{eq:network length}
\end{equation}
Note that the * superscripts in $D^*$ and $L^*$ indicate approximations of the exact number of hops $D$ and the exact end-to-end distance $L$. We use the analytical expressions for $\lambda$ derived in \cite{dawar2024quantum} for evaluating $D^*$ \eqref{eq:final_dstar}. However, the expression for $\lambda$ is based on the purification protocol by Bennett et.~al.~\cite{Bennett1996} (Equation \eqref{eq:purification eta eg 1}) applied in a nested sequence with one entanglement swap per iteration. \eqref{eq:final_dstar}, as well, is restricted to such a nested protocol.
The proceeding analysis could, however, be extended in future work by instead applying the more efficient protocol by Deutsch et.~al.~\cite{deutsch1996quantum} and generalizing to multiple entanglement swaps per iteration.

\section{Joint optimization: maximising network performance}
\label{sec:Joint optimization: maximising network performance}
We estimate the optimal end-to-end entanglement rate $R$ of a quantum network utilising a satellite assisted nested repeater protocol (Figure~\ref{fig:sat_repeater}). On the physical layer, we optimize with respect to mean photon-pair number $\mu$ of the SPDC source and the distance $L_0$ between adjacent ground stations. On the protocol layer, we optimize with respect to the target fidelity $F_t$ that the network should maintain using entanglement purification. 

While increasing $\mu$ enhances the elementary rate $R_0$ \eqref{eq:ultimate r0}, it also elevates the probability of multi-pair emissions. This leads to accidental coincidences that are indistinguishable from genuine entangled pairs, thus increasing the Quantum Bit Error Rate (QBER) \eqref{eq:qber main} and degrading the initial fidelity, $F_\text{init}$ \eqref{eq:f init main}. A lower initial fidelity forces the repeater protocol to execute more rounds of resource-intensive entanglement purification to reach a desired target fidelity, $F_t$. This, in turn, increases the time and entanglement resources required to establish an end-to-end link, and lowers $R$. If this time exceeds the memory coherence time, $T_2$, the protocol fails, rendering the network non-functional.

The choice of $\mu$ and $L_0$ determine the rate $R_0(\mu, L_0)$ and the initial fidelity $F_{\text{init}}(\mu, L_0)$. Once elementary entanglements with fidelities $F_{\text{init}}$ are generated, they need to be purified to fidelities $F_t$. These purified entanglements are extended using entanglement swapping, which lowers the fidelities again, this time from $F_t$ to a fidelity $F_0$. These are again purified back to the target fidelities $F_t$ and so on. This cycle of purification and swapping is repeated until an end-to-end entanglement of fidelity $F_t$ between the requested nodes is established (see Figure~\ref{fig:sat_repeater}). 

A low target fidelity $F_t$ might be easier to achieve in the beginning, but is harder to maintain, due to the larger amount of errors it accumulates for every swapping operation \eqref{eq:swap fidelity}. Likewise, a high $F_t$ is hard to maintain as well due to the concave nature of the purification map \eqref{eq:purification eta eg 1}, resulting in diminishing returns for purifications performed at already high fidelities. Indeed, as demonstrated in \cite{dawar2024quantum}, there is an optimal $F_t$ which minimises the resource scaling $\lambda$ of the network. Consequently, we pursue a joint optimisation of $(\mu, L_0, F_t)$.

We focus our analysis on ground nodes equipped with Nitrogen Vacancy (NV) centers, Silicon Vacancy (SiV) centers, and Neutral Atom qubits. These platforms are specifically selected due to their demonstrated capacity for absorptive quantum memories \cite{sunami2025scalable, nguyen2019integrated, ruf2021resonant}, which is essential for storing entangled photons received from satellite downlinks. The parameters that we consider for each of these platforms are given in Table \ref{tab: ground node params}.
\begin{table}[htbp]
    \centering
    \caption{Comparison of Platform-Specific Quantum Memory Parameters}
    \label{tab:platform_specific_params}
    \begin{tabular}{|c| c| c| c|}
    \hline
        \textbf{Parameter} & \textbf{SiV} & \textbf{NV} & \textbf{Atoms} \\
    \hline
        Read-out error $\epsilon_r$ &  $1 \times 10^{-4}$ \cite{bhaskar2020experimental}& $4 \times 10^{-4}$ \cite{zhang2021high} & 0.004 \cite{radnaev2024universal}\\
        Middle fixed-point $F_l$ & 0.5017 \cite{dawar2024quantum}& 0.5019 \cite{dawar2024quantum}& 0.5162 \cite{dawar2024quantum}\\
        Two-qubit gate infidelity $\epsilon_g$ & $5 \times 10^{-4}$ \cite{stas2022robust} & $3.5 \times 10^{-4}$ \cite{bartling2024universal} & 0.0025 \cite{evered2023high}\\
        Coherence time $T_2$ & 2.1 \cite{stas2022robust} & 75 \cite{bradley2019ten} & 42 \cite{barnes2022assembly}\\
        Efficiency of CAPS gate $\eta_{\text{CAPS}}$ & - & - & 0.75 \cite{sunami2025scalable}\\
        Detection efficiency $\eta_d$ & - & - & 0.9 \cite{sunami2025scalable}\\
        $\eta_{\text{CAPS}} \eta_d$ & 0.4 \cite{nguyen2019integrated} & 0.1 \cite{ruf2021resonant} & 0.675\cite{sunami2025scalable} \\
        Transition wavelength $\lambda$ (nm) & 738 \cite{hepp2014electronic}& 638 \cite{schirhagl2014nitrogen}& 780 \cite{steck2001rubidium}\\ 
        Spin-photon interface bandwidth $f_s$ (Hz) (Scenario A)\footnote{For Scenario A, we take $f_s$ to be the inverse of the excited state lifetime recorded in recent research. For Scenarios B and C, we consider $f_s$ to be $10^9$ for all the three platforms, matching the emission rate of $f$ of the SPDC source.} & $10^9$ \cite{nguyen2019quantum} & $10^8$ \cite{storteboom2015lifetime}& 3.34 $\times 10^8$ \cite{reiserer2015cavity}\\
    \hline
    \end{tabular}
    \label{tab: ground node params}
\end{table}

The most achievable use case for such a quantum network is considered to be quantum key distribution (QKD). Therefore, we consider quantum networks capable of achieving at least QKD, for which, ideally, the end-to-end entanglement fidelity should be $\geq95\%$, corresponding to a Quantum Bit Error Rate QBER $\approx 3\%$ \cite{paraiso2019modulator, pathak2023phase}. Furthermore, the source brightness $\mu$ needs to be $ < 0.5$ to ensure security against photon number splitting attacks \cite{schmitt2007experimental, peng2007experimental, lutkenhaus2000security}.

We consequently perform an optimization across the parameter space  $(\mu, F_t, L_0)$  for $\mu \in [2 \times 10^{-4}, 0.2]$, $F_t \in [0.95, 1]$ and $L_0 \in [500\text{ km}, L]$ where (see Appendix \ref{appendix:f} for the derivation):
\begin{equation}
    L = 2 R_E \left[\text{arccos}\left(\frac{R_E \cos(\phi_{\text{min}})}{R_E + h}\right) - \phi_{\text{min}}\right]
    \label{eq:lmax}
\end{equation}
where $R_E \approx 6371$ km is the radius of Earth and $\phi_{\text{min}}$ is the minimum angle of elevation where it is still practical to catch a signal from the satellite. We consider it to be $15^\circ$. The other system parameters are displayed in Table \ref{tab:unified_params_3_scenarios}.

\begin{table*}[t]
    \centering
    \caption{Parameters for satellite-to-ground link: Comparison of system parameters across three technology Scenarios. For detail derivation please see Appendix \protect\ref{app:error models}.}
    \label{tab:unified_params_3_scenarios}
    \begin{tabularx}{\linewidth}{|l| Y| c| c| c|}
    \hline
        \textbf{Symbol} & \textbf{Description} & \textbf{State-of-the-Art} & \textbf{Near-Term} & \textbf{Optimistic} \\
    \hline
    \multicolumn{5}{|c|}{\textbf{System Performance Parameters}} \\
    \hline
        $e_d$ \cite{de2023satellite}& Intrinsic system error rate & 0.01 & 0.01 & 0.01 \\
        $e_0$ \cite{de2023satellite}& Error rate of background counts & 0.5 & 0.5 & 0.5 \\
        $f$ \cite{ji2025global}& Emission rate of the source & $1 \times 10^9$ Hz & $1 \times 10^9$ Hz & $1 \times 10^9$ Hz \\
        $\mu$ \cite{schmitt2007experimental, peng2007experimental, lutkenhaus2000security}& Mean photon number per pulse & $\leq 0.5$ & $\leq 0.5$ & $\leq 0.5$ \\
        $N$ \cite{ji2025global}& Number of multiplexed modes & 1 & 20 & 100 \\
        $\eta_\text{demux}$ \cite{ji2025global}& Demultiplexing efficiency & 0.73 & 0.73 & 0.73 \\
    \hline
    \multicolumn{5}{|c|}{\textbf{Satellite and Link Geometry}} \\
    \hline
        $h$ \cite{mcdowell2020low}& Altitude of the LEO satellite (km) &\multicolumn{3}{c|}{ [500, 1000, 1500, 2000]} \\
        $L_0$ & Distance between ground stations (km) & \multicolumn{3}{c|}{ $L_0$ $\in$ [100\text{km}, L]} \\
        $R_E$ & Radius of the Earth (km) & \multicolumn{3}{c|}{ 6378} \\
    \hline
    \multicolumn{5}{|c|}{\textbf{Transmitter and Receiver Optics}} \\
    \hline
        $D_{TX}$ \cite{de2023satellite}& Diameter of transmitter telescope (cm) & 20  & 25 & 30 \\
        $D_{RX}$ \cite{de2023satellite}& Diameter of receiver telescope (cm) & 100  & 100  & 100 \\
        $\sigma_{\text{point}}$ \cite{lu2022micius}& Pointing error std. dev.  (µrad) & 2 & 1.5 & 1 \\
    \hline
    \multicolumn{5}{|c|}{\textbf{Channel and Noise Characteristics}} \\
    \hline
        $\eta_{\text{zenith}}$ \cite{liao2017long} & Atmospheric transmittance at zenith & 0.8 & 0.8 & 0.8 \\
        $\eta_{\text{cpl}}$ (Appendix \ref{app:coupling loss}) & Total internal \& coupling efficiency & 0.02 ($\sim$-17dB) & 0.1 ($\sim$-10dB) & 0.25 ($\sim$-6dB) \\
        $H_{\text{sky}}$ \cite{er2005background} & Spectral irradiance of sky (W m$^{-2}$ µm$^{-1}$ sr$^{-1}$)  & $1.5 \times 10^{-3}$  & $1.5 \times 10^{-3}$  & $1.5 \times 10^{-3}$  \\
        $\Omega_{\text{fov}} \eqref{eq:omegafov}$ & Field of view of the telescope (sr) & $8.87 \times 10^{-13}$ & $8.87 \times 10^{-13}$ & $8.87 \times 10^{-13}$ \\
        $\Delta\lambda$ \cite{ji2025global}& Spectral filter bandwidth (nm) &   1 & 1 & 1 \\
        $\Delta T$ \cite{ji2025global}& Coincidence time window (ns) & 1 & 1 & 1 \\
    \hline
    \end{tabularx}
\end{table*}

\subsection{Methods}
\label{sec:results}
We want to answer the following questions in the state of the art (Scenario A), near-term (5-10 years: Scenario B) and futuristic (10-15 years: Scenario C) time windows: 
\begin{enumerate}
    \item What are the end-to-end rates and fidelities that can we expect from a quantum network at: national (1,000 km), intercontinental (6,000 km) and planetary (15,000 km) scales?
    \item Should we use quantum repeaters for servicing networks at these scales or is direct satellite transmission to the end nodes good enough?
    \item What are the primary technological bottlenecks in the way of a high performing large-scale quantum internet?
\end{enumerate}
We consider space to ground entanglement transfer from LEO satellites stationed at altitudes ranging from 500 to 2,000 km. We take the number $N_m$ of photon modes, through spatial and spectral multiplexing, to be 1, 20 and 100 in Scenarios A, B and C respectively, which proportionally enhance the elementary rate $R_0$ \eqref{eq:ultimate r0}. We correspondingly multiplex the quantum memories on the ground, given the limited bandwidth $f_s$ of each memory. The pointing errors resulting from imperfect satellite tracking are taken to be 2, 1.5 and 1 $\mu$rad, which influence the overall transmission efficiency $\eta$ \eqref{eq: transmission efficiency main} inverse quadratically (see Equations \eqref{eq:diffraction loss} and \eqref{eq:corr_tx_gain}); the atmosphere to fiber coupling efficiencies on the ground: 0.02, 0.1 and 0.25, which proportionally enhance $\eta$; and the diameter of the transmitter telescopes on the satellites: 20, 25 and 30 cm, which enhances $\eta$ quadratically (see Equation \eqref{eq:diffraction loss}). These, along with the other parameters that we do not vary, have been listed in Table \ref{tab:unified_params_3_scenarios}.

In the following, we consider the end to end rates $R$ and fidelities $F_t$ with respect to the end-to-end distance $L$. The number of repeaters along a repeater chain of length $L$ is $L/L_0 - 1$, where $L_0$ is the distance between the adjacent OGS. The entanglement resources thereby consumed by each OGS-OGS link for distributing end to end entanglement is given by:
\begin{equation}
    T = \left(\frac{2}{\Tilde{P}_F}\right)^{m(\mu, L_0, F_t)}\left(2^{-s(\mu, L_0, F_t)}\frac{L}{L_0}\right)^{\lambda(F_t)-1}
\end{equation}
This is because each purification level consumes 2 entanglements along the same edge (OGS-OGS pair) and succeeds with probability $P_F$. Each swap, as well, consumes 2 entanglements, but along adjacent edges, doubling the link length with each nesting level. Once the fidelity reaches $F_t$ after initial purification and swapping, then the entanglement resources consumed by each entanglement link to distribute an end-to-end entanglement with fidelity $F_t$ are $\left(\frac{L}{\Tilde{L}_0}\right)^{\lambda - 1}$. Here, $\Tilde{L}_0$ is the length of an OGS-OGS link after initial swapping along $s$ levels: $\Tilde{L}_0 = 2^s L_0$. Note that we take $\left(2^{-s(\mu, L_0, F_t)}\frac{L}{L_0}\right) \rightarrow \text{max}\left[\left(2^{-s(\mu, L_0, F_t)}\frac{L}{L_0}\right), 1\right]$. This is because the resource scaling $T$ becomes unphysical for $\left(2^{-s(\mu, L_0, F_t)}\frac{L}{L_0}\right) < 1$.

Since $R_0$ is the entanglement rate between adjacent OGS, the end to end rate $R$:
\begin{equation}
    R(\mu, L_0, F_t) = \frac{R_0(\mu, L_0)}{T}
    \label{eq: rate equation}
\end{equation}

In the following, $R$ is maximized with respect to ($\mu, L_0, F_t$) with the conditions that $D^*(\mu, L_0, F_t) \geq 1$, $D^* \times L_0 \geq L$ (see Equations \eqref{eq:final_dstar} and \eqref{eq:network length}).
$D^*<1$ indicates networks that do not support even two nodes and $D^* \times L_0 < L$, those networks that cannot distribute end-to-end entanglement with fidelity $F_t$ across a distance of $L$ km.

\section{Results}
\subsection{Scenario A}
\begin{figure}[ht]
    \centering
    \includegraphics[width=\textwidth]{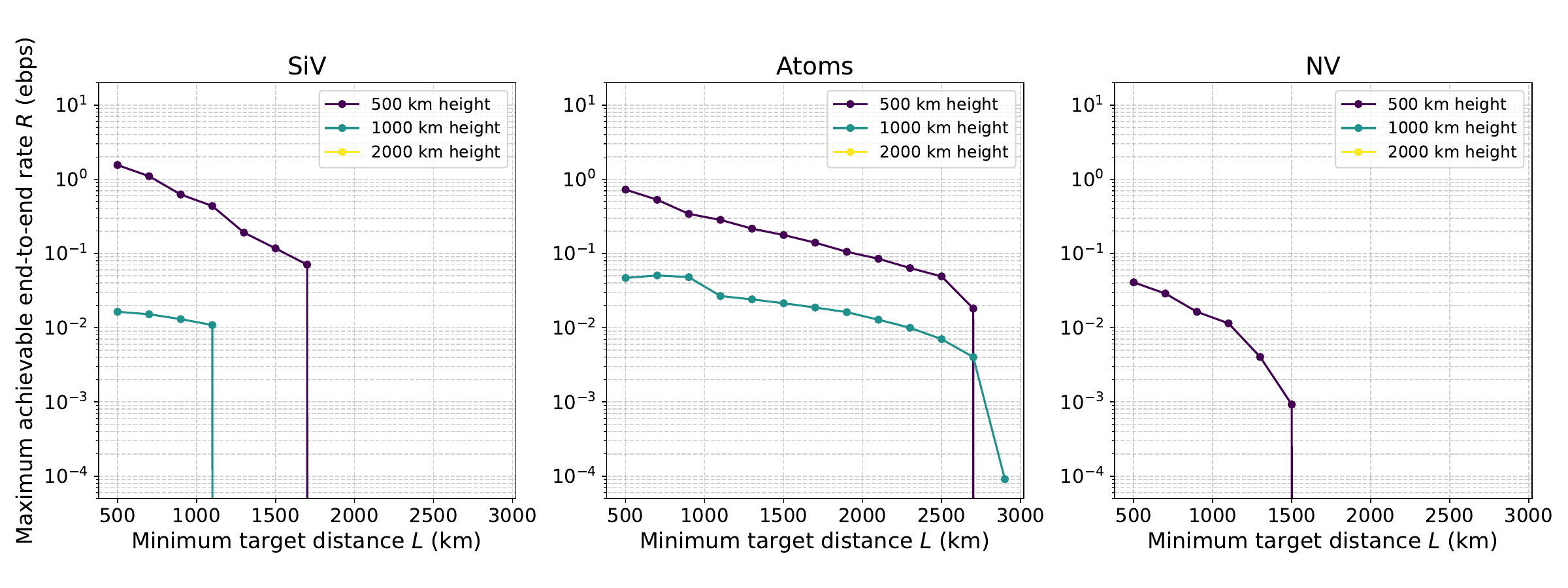}
    \caption{End-to-end rates using state of the art space technology (Scenario A): plotted across three quantum repeater platforms -- NV and SiV centers and Atoms -- and different satellite altitudes: from 500 to 2,000 km.}
    \label{fig:max_rate_vs_min_distance_pessimistic}
\end{figure} 
In Figure \ref{fig:max_rate_vs_min_distance_pessimistic}, we plot the maximum achievable end-to-end rate $R$ \eqref{eq: rate equation} vs the distance $L$ between the end nodes in the state of the art Scenario A. At $\approx 1000$ km, the rates are $\mathcal{O}(10^{-1})$ for SiV centers and Atoms. For NV centers, they are $\mathcal{O}(10^{-2})$. This disparity is because of the differences in the quantum memory absorption efficiencies $\eta_\text{CAPS} \eta_d$ and the spin-photon interface bandwidths $f_s$, which proportionally influence the elementary entanglement rate $R_0$ \eqref{eq:ultimate r0}. Notably, NV centers have the lowest $\eta_\text{CAPS} \eta_d = 0.1$ and the smallest $f_s=10^8$ Hz (see Table \ref{tab:platform_specific_params}).

Networks in Scenario A appear to be functional only with low altitude satellites. This is because lower orbits minimize the slant distance $d_\text{min}$ \eqref{eq:slant distance} between the satellite and ground station, minimizing diffraction loss \eqref{eq:diffraction loss}. Given the high pointing error ($2 \text{ } \mu\text{rad}$) and low coupling efficiency ($0.02$) in Scenario A, the quadratic increase in diffraction loss \eqref{eq:diffraction loss} at higher altitudes ($>1000$ km) (see Appendix \ref{app:error models}) reduces the elementary entanglement generation rate $R_0$ \eqref{eq:ultimate r0} to impractical levels, bringing $D^*\times L_0 < L$ \eqref{eq:final_dstar} for most configurations.

A combination of the absorption efficiency $\eta_\text{CAPS}$, interface bandwidth $f_s$ and the coherence time $T_2$ determine the maximum reach of the network. The former two influence the elementary rate $R_0$ proportionally \eqref{eq:ultimate r0}. $D^*$ scales with $R_0$ and $T_2$ as: $D^* \sim (R_0 T_2)^{1/(\lambda -1)}$ \eqref{eq:final_dstar}. Atoms reach the farthest because of their modest $f_s = 3.34 \times 10^8$ Hz, long coherence time (42 s) and high absorption efficiency $\eta_\text{CAPS} \eta_d \approx 0.675$ (see Table \ref{tab: ground node params}).
\begin{figure}[htbp]
    \centering
    \includegraphics[width=\textwidth]{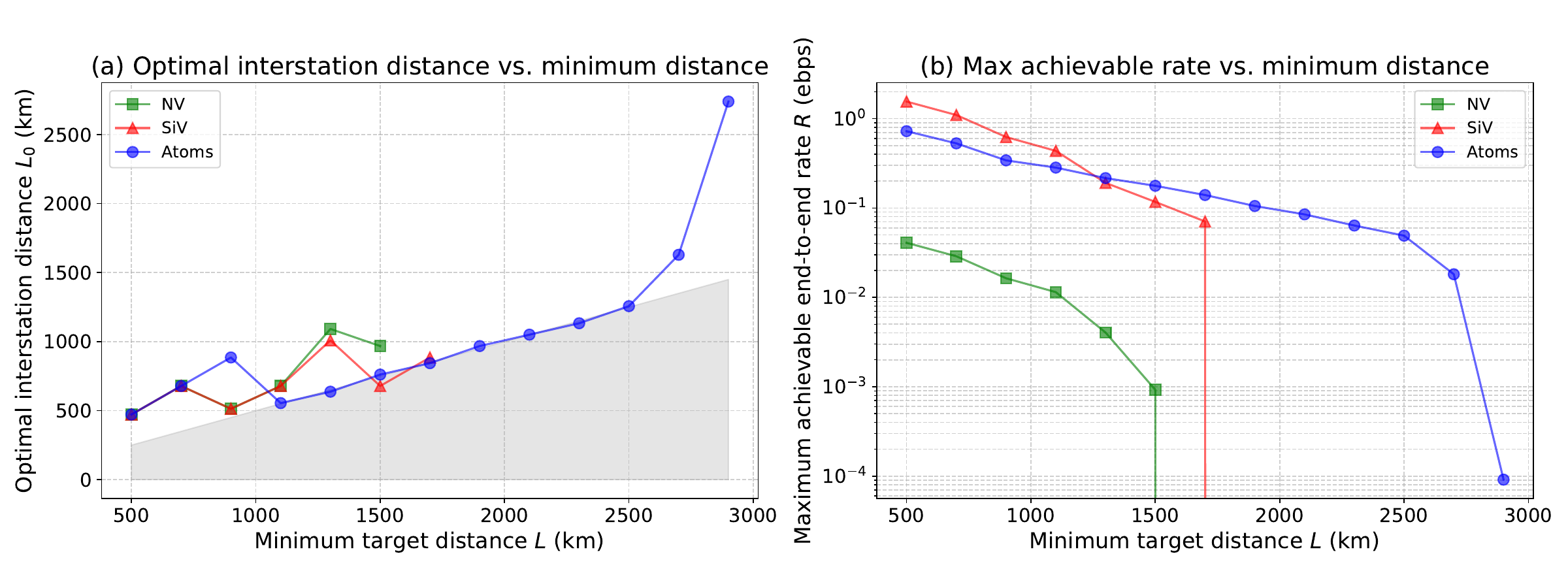}
    \caption{(a) The inter-station distance $L_0$ that maximizes the end-to-end rate $R$. The area under the line $L=L_0$, where repeaters are useful, has been shaded (b) maximum achievable end-to-end rate $R$, accounting for all satellite altitudes plotted against the distance between the end nodes for Neutral Atoms, Silicon Vacancy (SiV), and Nitrogen Vacancy (NV) platforms.}
    \label{fig:opt_L0_vs_min_distance_pessimistic}
\end{figure}
In Figure \ref{fig:opt_L0_vs_min_distance_pessimistic}(a), we plot the interstation distance $L_0$ that maximises the rate $R$ at a given distance $L$. For Atoms, until about 1,000 km, $L_0 = L$, meaning that a direct transmission from the satellites to the end ground nodes, without performing swapping or purification on intermediate repeater nodes, is optimal. Beyond that, until 2,500 km, $L_0 = L/2$, implying that it is optimal to have one intermediate repeater node. Consequently, the area where $L_0 < L/2$, is shaded to demarcate the regime where quantum repeaters unambiguously provide utility.
For NV and SiV centers, direct transmission is optimal until 750 km. Beyond that, both fail to scale efficiently: for NV centers, this is because of their low interface bandwidth $f_s = 10^8$ Hz, while for SiV centers, the issue is their relatively low coherence time $T_2 = 2.1$ s. 
 
In Figure \ref{fig:opt_L0_vs_min_distance_pessimistic}(b), we plot the maximum rate across all satellite altitudes for the three platforms. SiV centers start at the highest rates because of their large spin photon interface bandwidth $f_s = 10^9$ Hz and decent absorption efficiency $\eta_\text{CAPS} \eta_d = 0.4$. Atoms, with their longer coherence time $T_2 = 42$ s, however, manage to scale better with distance and reach further.

\subsection{Scenario B}
\begin{figure}[htbp]
    \centering
    \includegraphics[width=\textwidth]{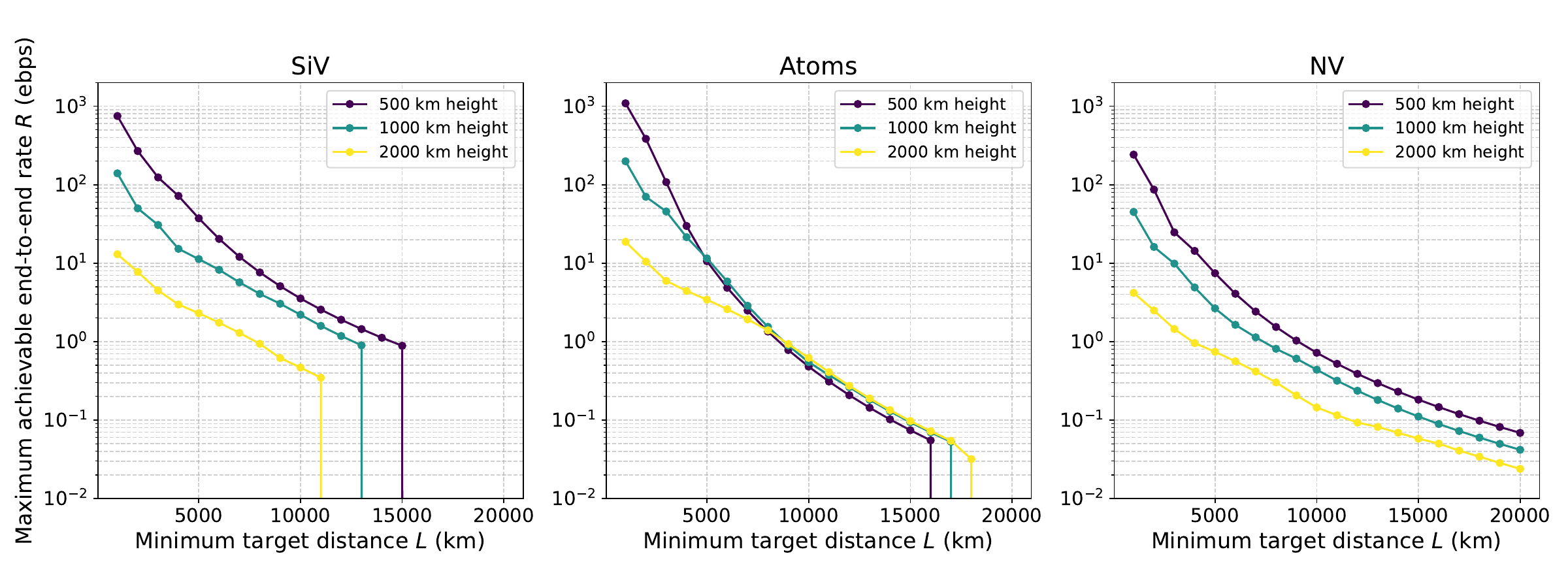}
    \caption{End-to-end rates using near-future space technology (Scenario B): plotted across three quantum repeater platforms -- NV and SiV centers and Atoms -- and different satellite altitudes: from 500 to 2,000 km.}
    \label{fig:max_rate_vs_min_distance_realistic}
\end{figure}

\begin{figure}[htbp]
    \centering
    \includegraphics[width=\textwidth]{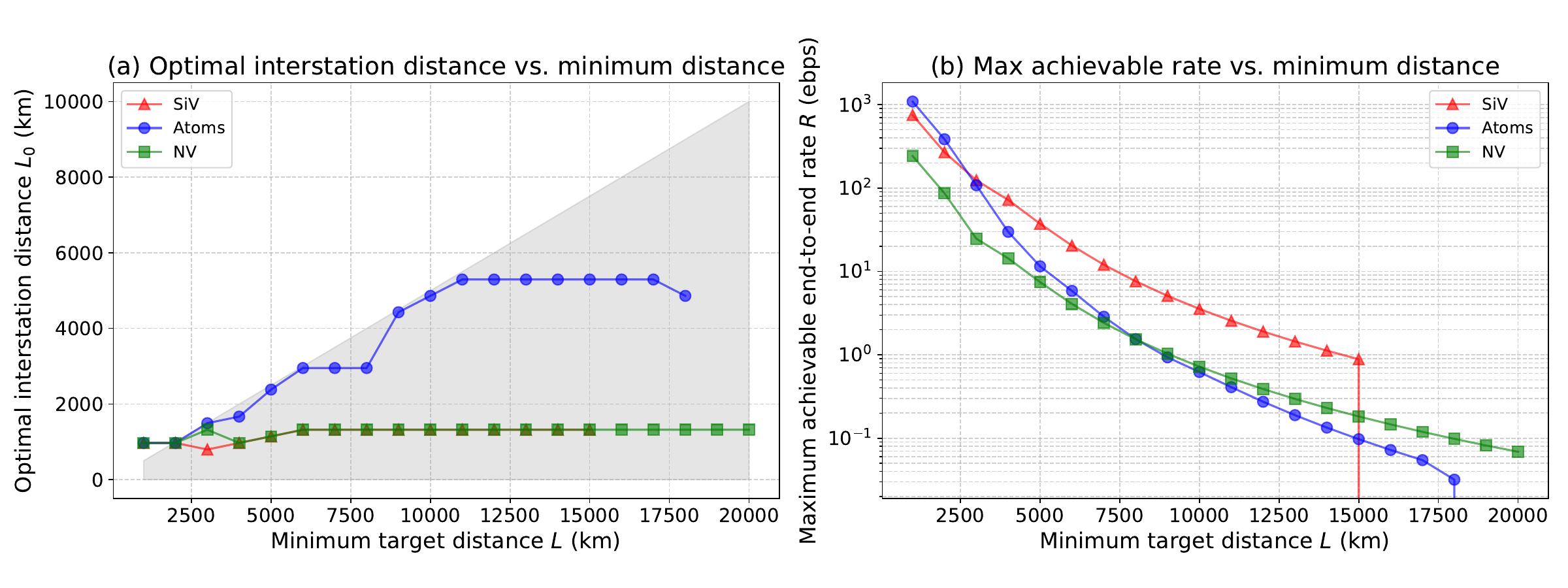}
    \caption{(a) The inter-station distance $L_0$ that maximizes the end-to-end rate $R$. The area under the line $L=L_0$, where repeaters are useful, has been shaded (b) maximum achievable end-to-end rate $R$, accounting for all satellite altitudes plotted against the distance between the end nodes for Neutral Atoms, Silicon Vacancy (SiV), and Nitrogen Vacancy (NV) platforms.}
    \label{fig:opt_L0_vs_min_distance_realistic}
\end{figure}

In Scenario B, the integration of near-future space technologies, specifically the introduction of multiplexing ($N_m=20$), improved coupling efficiency ($\eta_\text{cpl} = 0.1$), and lower pointing error (1.5 $\mu$rad), yields significant performance enhancements. Moreover, we consider that the interface bandwidth $f_s$ now matches the source bandwidth $f_s = f = 10^9$ Hz for all the three platforms.

As shown in Figure~\ref{fig:max_rate_vs_min_distance_pessimistic}, achievable entanglement rates increase by approximately three orders of magnitude compared to Scenario A, extending the viable network range already to 20,000 km for NV centers, 18,000 km for Atoms, and 15,000 km for SiV centers. Low altitude satellites generally perform better because of lower diffraction losses due to the reduced slant distance between the satellites and the ground nodes. In Atoms, the satellite altitude curves intersect at 8,000 km because the low altitude scenario requires more repeater nodes. Since Atoms have a relatively larger resource scaling exponent $\lambda$, they are not able to use repeaters as efficiently as the diamond color center platforms. 

Figure~\ref{fig:opt_L0_vs_min_distance_realistic}(a) demonstrates that quantum repeaters become more significant compared to Scenario A, with improvements in the (complementary) space technology. NV and SiV center networks are optimized for repeaters placed every 1,200 km. Atoms, on the other hand stay roughly on the $L_0 = L/2$ line, inclining towards one repeater networks. 

In Figure \ref{fig:opt_L0_vs_min_distance_realistic}(b), we plot the maximum rate for all the platforms among all the satellite altitudes. Atoms start at the highest rates, followed closely by SiV, and then NV centers. This is because of their respective absorption efficiencies: $\eta_\text{CAPS}\eta_d = 0.675, 0.4$ and 0.1 respectively. The SiV and NV center curve slopes are similar, and more forgiving than the one for Atoms, indicating their more efficient use of repeaters (smaller $\lambda$). At long distances, therefore, a smaller $\lambda$ becomes significant. \cite{dawar2024quantum} showed that $\lambda$ depends largely on the two-qubit gate infidelity $\epsilon_g$. SiV centers, however, are not able to extend beyond 15,000 km despite providing the highest rates until then, because of their smaller coherence time $T_2=2.1$ s. Atoms do not go beyond 18,000 km mainly because of their large $\lambda$, which restricts $D^* \times L_0 \leq 18,000$ km \eqref{eq:network length}. 

\subsection{Scenario C} 
\begin{figure}[htbp]
    \centering
    \includegraphics[width=\textwidth]{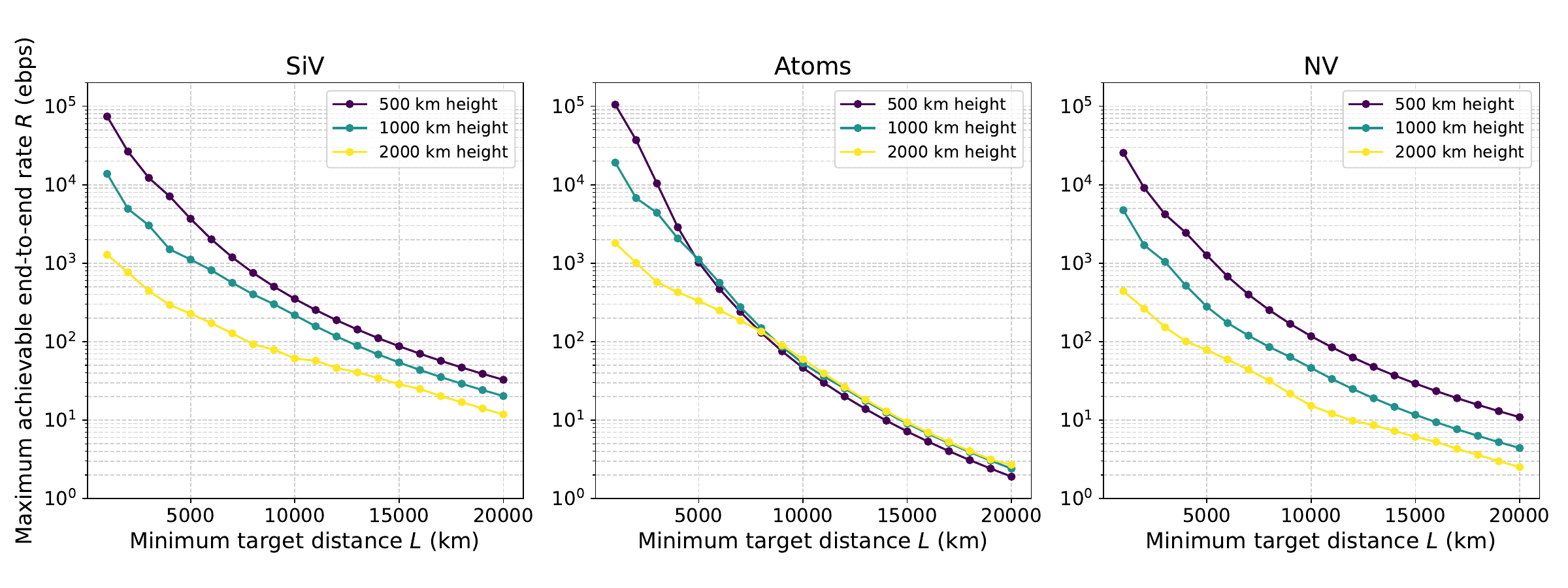}
    \caption{End-to-end rates using futuristic space technology (Scenario C): plotted across three quantum repeater platforms -- NV and SiV centers and Atoms -- and different satellite altitudes: from 500 to 2,000 km.}
    \label{fig:max_rate_vs_min_distance_optimistic}
\end{figure}

\begin{figure}[htbp]
    \centering
    \includegraphics[width=\textwidth]{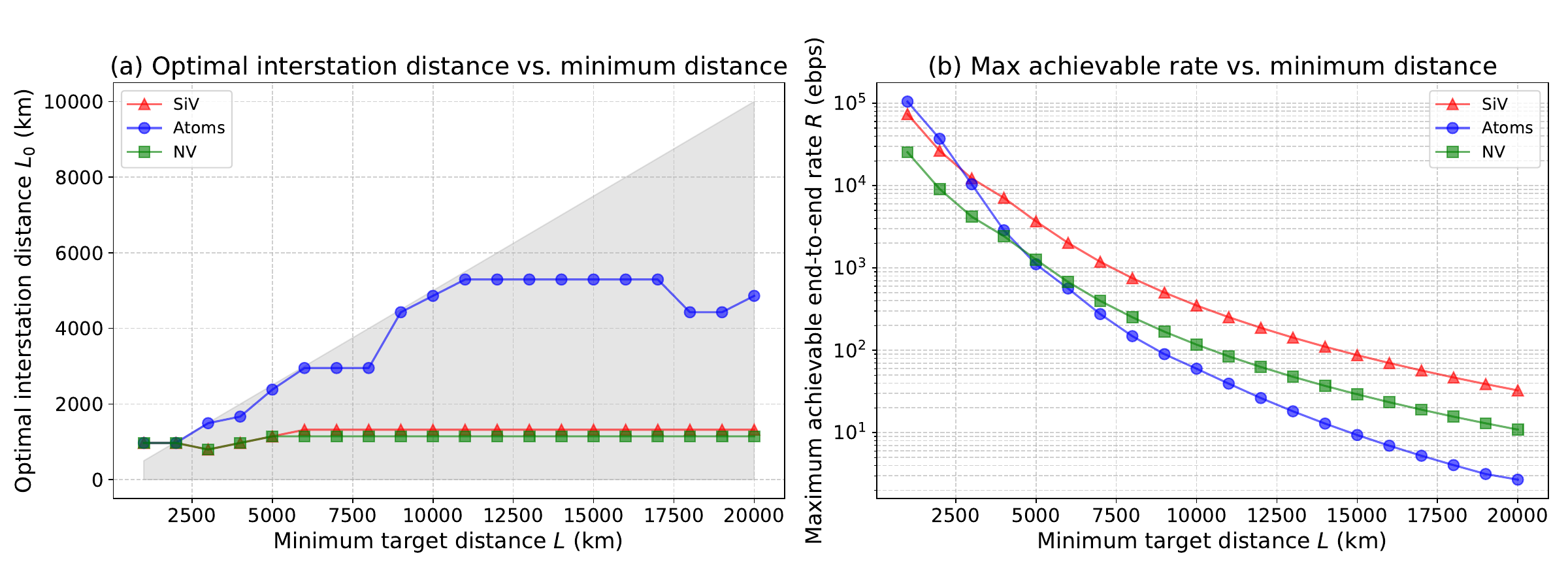}
    \caption{(a) The inter-station distance $L_0$ that maximizes the end-to-end rate $R$. The area under the line $L=L_0$, where repeaters are useful, has been shaded (b) maximum achievable end-to-end rate $R$, accounting for all satellite altitudes plotted against the distance between the end nodes for Neutral Atoms, Silicon Vacancy (SiV), and Nitrogen Vacancy (NV) platforms.}
    \label{fig:opt_L0_vs_min_distance_optimistic}
\end{figure}
Finally, in Figure \ref{fig:max_rate_vs_min_distance_optimistic}, we analyze the performance enabled by futuristic space technology. Defined by generous multiplexing ($N_m=100$), low pointing errors ($1~\mu\text{rad}$), and optimized coupling ($\eta_\text{cpl}=0.25$), this scenario enhances the performance dramatically: ranging from about a hundred kilohertz ($10^5$ ebps) at national scales ($\approx 1,000$ km) to sustained rates of $\approx 100$ ebps at $15,000$ km. Atoms start at the highest rates because of their large absorption efficiencies, followed closely by SiV, then NV centers. Here, as well, like in Scenarios A and B, low altitude satellites perform better because of their lower diffraction losses. The satellite altitude curves for Atoms, however, intersect at $>7,500$ km because of their larger $\lambda$, meaning that they are not able to use repeaters efficiently, making it better to have larger hop lengths $L_0$ enabled by the high altitude satellites, rather than more repeaters.

Figure \ref{fig:opt_L0_vs_min_distance_optimistic}(a) confirms that repeaters become even more significant here, allowing all platforms to scale to 20,000 km, which is already the farthest distance between any two points along a geodesic on Earth's surface. Because of their small $\lambda$, NV and SiV centers efficiently use repeaters placed every 1,200 - 1,500 km across the globe. NV centers, on the other hand, favor one repeater networks until about 11,000 km. At 20,000 km, however, they optimize for $L_0/L = 1/4$: 3 repeaters. This is reflected in Figure \ref{fig:opt_L0_vs_min_distance_optimistic}(b), where the rate curves follow the same trajectory as in Scenario B. SiV maintain the highest rates at long distances because of their low $\lambda$ and decent absorption efficiency $\eta_\text{CAPS}\eta_d=0.4$.

\subsection{Rate vs fidelity}
\begin{figure}[htbp]
    \centering
    \includegraphics[width=\textwidth]{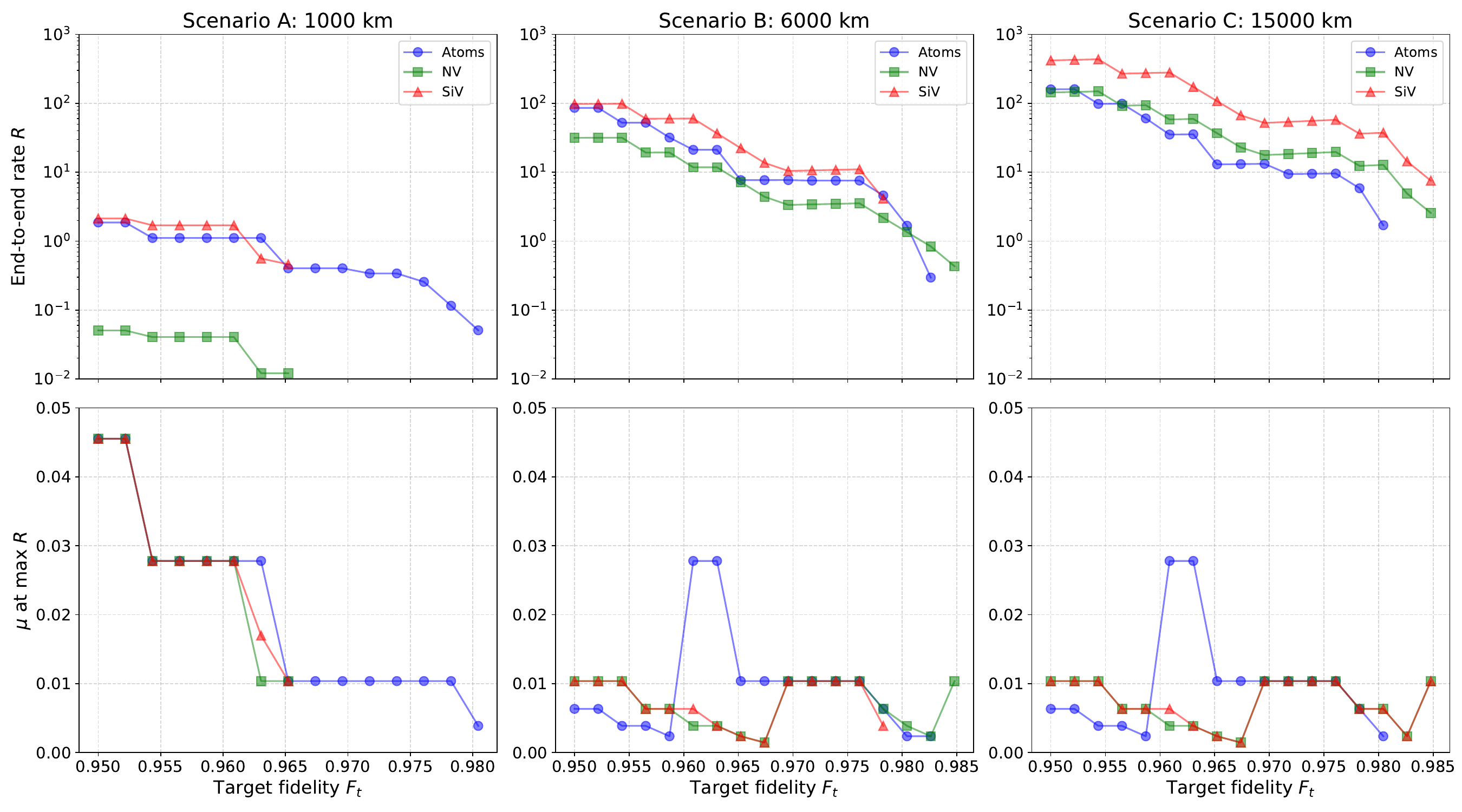}
    \caption{Rate (top) and $\mu$ (bottom) vs fidelity tradeoff for the three scenarios. While a low $\mu$ enables high fidelities, it comes at the cost of a low rate $R_0$ \eqref{eq:ultimate r0}.}
    \label{fig:scenarios_comparison}
\end{figure}
We have provided a comprehensive performance evaluation for a large scale quantum repeater network, evaluating architectures across national, intercontinental, and planetary scales. Our analysis considers three distinct Scenarios (A, B, and C) based on the anticipated maturity of space-segment technologies.
We discussed the optimal rates at fidelities $\geq95\%$, recommended for QKD. Other applications like distributed or blind quantum computing, however, might demand even higher fidelities. In Figure \ref{fig:scenarios_comparison} (top), we plot the end-to-end rates $R$ vs end-to-end fidelities $F_t$ at distances of 1,000, 6,000 and 15,000 km respectively for scenarios A, B and C. Generally, high fidelities mean lower rates, but sometimes the rates peak at intermediate fidelities because high fidelities minimize $\lambda$, as we discussed in Section \ref{sec:Joint optimization: maximising network performance}.

In Scenario A, NV and SiV centers are not able to offer fidelities exceeding $F_t=0.965$. For NV centers, this is because of their low interface bandwidth $f_s = 10^8$ Hz and low absorption efficiency $\eta_\text{CAPS} \eta_d = 0.1$. Both these factors proportionally affect the elementary rate $R_0$ \eqref{eq:ultimate r0}, explaining why NV centers have lower end-to-end rates $R$ \eqref{eq: rate equation}. This, in turn influences the maximum number of hops $D^*$ \eqref{eq:final_dstar}, bringing many configurations to $D^*<1$. SiV centers are not able to exceed $F_t=0.965$ because of their low coherence time $T_2=2.1$ s, which influences $D^*$, rendering many configurations non-functional: $D^*<1$.

In Scenario B, SiV centers maintain the highest rates until $F_t = 0.976$. This is because of their small $\lambda$, allowing them to use repeaters more efficiently and decent absorption efficiency $\eta_\text{CAPS} \eta_d = 0.4$. Beyond $F_t = 0.976$, their coherence time $T_2=2.1$ s runs out. The NV center curve has the same shape until $F_t = 0.976$, because they have a similar exponent $\lambda$ as that for SiV. However, their absorption efficiency is lower at $\eta_\text{CAPS} \eta_d = 0.1$, explaining the constant offset between the two curves. However, they are able to go up to $F_t =0.985$ because of their longer coherence time $T_2 = 42$ s. 

In Scenario C, the coherence time does not remain a limitation anymore for SiV, allowing them to offer the highest rates at all fidelities. Their curve is shaped identically to the one for NV centers, which have a similar $\lambda$, but is offset higher because of their larger absorption efficiency. Atoms with their larger absorption efficiency $\eta_\text{CAPS} \eta_d = 0.675$, start at higher rates than NV centers at $F_t=0.95$, but drop off quicker because of their larger $\lambda$.

Therefore, once space technology gets mature enough, as in Scenario C, the coherence time does not matter anymore, as long as it is a few seconds. The most important considerations then are, the exponent $\lambda$, which decides the shape of the curve and largely depends on the two-qubit gate infidelity, and the absorption efficiency, which decides the starting point of the curve. 

In Figure \ref{fig:scenarios_comparison} (bottom), we plot the corresponding mean photon-pair numbers that optimize the rates. For Scenario A, the trend is clear: the higher the fidelities $F_t$ that we want to maintain, the lower the $\mu$ should be. $\mu$ goes from 0.045, corresponding to an initial fidelity $F_\text{init} \approx 0.97$ to $\mu \approx 10^{-3}$, $F_\text{init} \approx 0.985$. Generally, the optimization suggests that $F_\text{init} > F_t$ to allow for the number of initial swaps $s$ to be $> 0$ (see Equations \eqref{eq:final_dstar} and \eqref{eq: rate equation}). 

Scenarios B and C exhibit almost identical trends. Initially, the rate is optimized for low $\mu$, allowing for two initial swaps to go from $F_\text{init}$ to $F_t$, with a downward trajectory continuing until around $F_t \approx 0.96$ for Atoms and $F_t \approx 0.966$ for NV and SiV centers. Thereafter, there is a jump to a larger $\mu$ indicating that now, one initial swap is more optimal than two swaps, since larger $\mu$ allow for larger elementary rates \eqref{eq:ultimate r0}. After this jump, the downward trend continues again. For NV centers, there is a jump again at $F_t = 0.985$ to $\mu = 0.01$, corresponding to $F_\text{init}\approx 0.985$ as well, indicating that at this point, it is best to have no initial swaps $s$ or purifications $m$.

\section{Discussion and conclusion}
\label{sec:discussion and conclusion}

We estimate the performance of near-term terrestrial quantum-repeater networks using Low Earth Orbit (LEO) satellites as entanglement sources. The analysis was performed for quantum repeaters, performing gate-based entanglement swapping and purification using Neutral Atoms, Silicon Vacancy (SiV) centers and Nitrogen Vacancy (NV) centers based quantum computing technologies. We consider three levels of the anticipated technological maturity of space systems technology: 

Scenario A is characterized by state of the art space technology, limited by a large pointing error ($\sigma_\text{rad} = 2 \text{ }\mu$rad), low atmosphere to fiber coupling efficiency $\eta_\text{cpl} = 0.02$ on the ground, no access to multiplexing and small transmitter telescopes on the satellites ($D_\text{Tx} = 20$ cm). Quantum networks in this scenario are limited to national scales $L<3,000$ km, with distribution rates $\mathcal{O}(10^{-1})$ ebps at end-to-end fidelities near $95\%$. Quantum repeaters provide limited utility, restricting the networks to repeater-less or single repeater architectures. Neutral Atom networks, with their long coherence time $T_2 = 42$ s, are able to service the largest distances up to $L \approx 3,000$ km. SiV centers, with their large interface bandwidth $f_s = 10^9$ Hz and decent absorption efficiency $\eta_\text{CAPS} \eta_d = 0.4$, provide the largest rates up to $L=1,200$ km, but are not able to extend beyond $L=1,700$ km because of their lower coherence time $T_2=2.1s$. NV centers, with their lower interface bandwidth ($f_s = 10^8$ Hz) and low absorption efficiency ($\eta_\text{CAPS} \eta_d = 0.1$) are restricted to the lowest rates ($R<10^{-1}$) and shortest distances ($L<1,500$ km). 

Advancements anticipated within a $\approx 5$-year horizon facilitate the transition to Scenario B. This regime supports around a kilohertz-rate entanglement distribution at national scales $\approx 1,000$~km and maintains $\approx 10$ ebps at intercontinental ranges ($\approx 6,000$~km). These enhancements are primarily driven by an increase in the number of multiplexed modes to $N_m = 20$, a reduction in satellite pointing error to $1.5~\mu\text{rad}$, and an improvement in free space to fibre coupling efficiency to $\eta_\text{cpl} = 0.1$. The impact of these improved parameters is a substantial reduction in the total link loss budget, which elevates the elementary entanglement generation rate $R_0$. Crucially, our optimization reveals the emergence of useful $>1$ repeater architectures for NV and SiV centers. Their low resource scaling exponent $\lambda$ allows these platforms to efficiently use repeater nodes over large distances. This enables larger rates compared to those with Atoms beyond certain distances: specifically, SiV centers start outperforming Atoms beyond 3,000 km, and NV centers start outperforming them beyond 8,000 km. At an intercontinental distance (6,000 km), however, SiV centers fail to distribute entanglements at fidelities exceeding 97.7\% because of their lower coherence time of 2.1 s. Atoms and NV centers are able to go up to 98.3\% and 98.5\% respectively with their coherence times of 42 s and 75 s.   

Improving space technology even further, as we suspect, might be achievable in $\approx 15$ years (Scenario C), global connectivity with $L=20,000$~km becomes possible. This regime is enabled by high-density multiplexing ($N_m = 100$), precise pointing ($1~\mu\text{rad}$), and optimized coupling efficiency reaching $0.25$. The impact of these advancements is $\approx 100$ kHz entanglement distribution rates at national scales and multiple kilohertz at intercontinental distances. For Atoms, $>1$ repeater architectures become optimal beyond 15,000 km. Although they provide the largest rates at small distances, they are overtaken by SiV centers at $3,000$ km, and by NV centers at 5,000 km. SiV and NV centers scale better with distance because of their lower exponent $\lambda$, allowing them to use repeaters more efficiently. SiV, however, start at higher rates because of their larger absorption efficiency. In terms of the end-to-end fidelities as well, now SiV centers are able to deliver higher rates at fidelities up to 98.5\% without their 2.1 s coherence time becoming a limitation. They are able to outperform NV centers and Atoms, both, in terms of end-to-end rates at large distances, and at large fidelities, despite them supporting significantly longer coherence times of 75 s and 42 s.

Therefore, with advanced space technology, the two most decisive factors for quantum network performance are the resource scaling exponent $\lambda$ and the absorption efficiency $\eta_\text{CAPS}\eta_d$. $\lambda$ measures how efficiently quantum repeaters are used: the lower, the better. The major contribution to $\lambda$ comes from the two-qubit gate infidelity. Improving the two-qubit gate fidelity should thence yield drastic improvements in the end-to-end rates and fidelities of quantum repeater networks. $\eta_\text{CAPS}\eta_d$ is the efficiency with which entangled photons arriving from space are absorbed into the quantum memories on the ground. 

Realizing a performant global quantum repeater network requires synergetic enhancements across space and quantum hardware technologies. Efforts into improving space technologies should focus on enabling efficient multiplexing of photon modes, improving atmosphere to fiber coupling efficiencies, reducing pointing errors and fitting larger transmitter telescopes on LEO satellites. On the quantum hardware front, the biggest impact will come from improving two-qubit gate fidelities and enhancing the absorption efficiencies of quantum memories.  

Future work could consider more efficient repeater protocols, which, for example, allow for more than one entanglement swapping per nesting level to optimize the end-to-end rates and fidelities even further. Dynamic weather and daylight conditions could be incorporated. Morover, dynamic satellite orbits could be taken into account. 

\section{Acknowledgments}
This research was cofinanced by Airbus. R.R. acknowledges support from the Cluster of Excellence “Advanced Imaging of Matter” of the Deutsche Forschungsgemeinschaft, EXC 2056, Project No. 390715994 and Bundesministerium für Bildung und Forschung via project QuantumHiFi (Grant No. 16KIS1592K), Forschung Agil. The project is cofinanced by ERDF of the European Union and by “Fonds of the Hamburg Ministry of Science, Research, Equalities and Districts.”

\bibliographystyle{naturemag}
\bibliography{bibl}

\newpage

\appendix

\section{Theoretical framework for quantum repeater chains}
\label{app:Theoretical Framework for Repeater-Assisted Networks}
To analyze the performance of a large-scale quantum repeater network, we adopt the resource estimation framework for first-generation quantum repeaters developed by Dawar et.~al.~\cite{dawar2024quantum}. This framework provides an analytical link between the microscopic errors in a quantum processor and the macroscopic performance of the resultant network leveraging a nested repeater protocol.

\subsection{Nested repeater protocol}
\label{sec:nested repeater protocol}
The model considers a linear chain of quantum repeater nodes. The protocol for establishing a long-distance entangled pair operates through a nested, recursive procedure. 

\begin{enumerate}
    \item \textbf{Elementary Link Generation:} The process begins by establishing entangled pairs between adjacent nodes. These are the elementary links, each with an initial length $L_{0}$ and an initial fidelity $F_\text{init}$.
    \item \textbf{Entanglement Purification:} Fidelity can be enhanced using entanglement purification. In particular, multiple rounds could be used to purify entanglements from fidelities $F_\text{init}$ to desired fidelities $F_t$. 
    In this work, we specifically consider the CNOT-based scheme by Bennett et.~al.~\cite{Bennett1996}, which uses multiple links with fidelities $F$ to distill a single link of higher fidelity $F'$ \cite{dawar2024quantum}:
    \begin{equation}
    \begin{matrix}
        F'(F,\eta, \epsilon_g) = \frac{\left[ F^2 + \left(\frac{1-F}{3}\right)^2 \right] \left[\eta_\text{ro}^2 + (1 - \eta_\text{ro})^2\right] + \left[F\left(\frac{1-F}{3}\right) + \left(\frac{1-F}{3}\right)^2\right]2\eta_\text{ro}(1 - \eta_\text{ro}) + 2\left(\frac{2\epsilon_g - \epsilon_g^2}{(1-\epsilon_g)^2}\right)\left[p_z F\left(\frac{1-F}{3}\right) + (p_x + p_y) \left(\frac{1-F}{3}\right)^2\right]}{\frac{1}{(1-\epsilon_g)^2}\left( \left[ F^2 + 2F\left(\frac{1 - F}{3}\right) + \frac{5}{9}\left(1 - F\right)^2 \right] \left[\eta_\text{ro}^2 + (1 - \eta_\text{ro})^2 \right] + \left[F\left(\frac{1-F}{3}\right) + \left(\frac{1-F}{3}\right)^2\right]8\eta_\text{ro}(1 - \eta_\text{ro}) \right)}
    \label{eq:purification eta eg 1}
    \end{matrix}
    \end{equation}
    where $\epsilon_g$ is the probability of a gate error, $p_{x/y/z}$ is the probability, specifically for a $\sigma_{x/y/z}$ quantum computing gate error, conditioned on $\epsilon_g$ and $\eta_\text{ro}$ is the read-out error \cite{dawar2024quantum}.
    In particular, $(1-\epsilon_g)^2$ times the denominator is the purification acceptance probability $P_F$ and $(1-\epsilon_g)^2$ times the numerator is the purification success probability conditioned on acceptance \cite{dawar2024quantum}.
    \item \textbf{Entanglement Swapping:} To extend the range, entanglement swapping is performed at alternate nodes. A Bell State Measurement (BSM) on two qubits from adjacent links effectively connects them, creating a new entangled pair spanning twice the distance. However, this process degrades the fidelity to a value $F_0$. For a Werner state with fidelity $F_{\text{init}}$, the fidelity after swapping is \cite{briegel1998quantum}:
    \begin{equation}
    F_0 = \frac{1}{4}\left\{1+3\left(\frac{4\eta_s^2-1}{3}\right)\left(\frac{4F_{\text{init}}-1}{3}\right)^2\right\}
    \label{eq:swap fidelity}
    \end{equation}
    where $\eta_s$ is the effective readout efficiency of the BSM, which, as in \cite{dawar2024quantum}, we take to be the same as the one in purification: $\eta_s = \eta_\text{ro}$
    \item \textbf{Nesting:} This sequence of swapping and purification constitutes one nesting level. The procedure is repeated recursively, doubling the link length at each level, until a single high-fidelity entangled pair is established between the end-users.
\end{enumerate}

\section{Loss modelling: Optical Antenna Gain}
\label{appendix:Loss modelling: Optical Antenna Gain}
In the realm of quantum optical communication from space, the precise characterization of optical antenna gain is of great importance. This appendix draws upon the work of Klein and Degnan \cite{klein1974optical} to justify the use of their model for centrally obscured optical antennas in the context of quantum optical communication systems. Quantum optical communication, often relies on laser systems employing telescopes as optical antennas, which frequently have central obscurations due to their design, such as the secondary mirror in Cassegrain or Gregorian telescopes. The presence of these obscurations can significantly impact the antenna gain and beam divergence, factors that are critical for the successful implementation of quantum communication protocols over long distances. The work of \cite{klein1974optical} provides a comprehensive analysis of the gain patterns for centrally obscured optical antennas, both in the near- and far-field. Their derivation of a simple polynomial equation for matching the incident source distribution to a general antenna configuration allows for the optimization of on-axis gain. This is particularly relevant for quantum optical communication, where maximizing the on-axis gain can enhance the budget link and improve the overall performance of the quantum communication link. The use of this model is favored over traditional Gaussian beam propagation models for several reasons:
\begin{enumerate}
    \item \textbf{Central Obscuration:} The model explicitly accounts for central obscurations, which are common in space-based optical systems. Gaussian beam propagation models do not inherently consider these obscurations, leading to potential inaccuracies in gain and divergence calculations.
    \item \textbf{Far-Field Performance:} The model provides detailed far-field gain patterns, which are crucial for assessing the performance of quantum communication links over large distances. This includes the evaluation of beam divergence and the impact of pointing errors.
    \item \textbf{Design Flexibility:} The polynomial equation derived by Klein and Degnan allows for the optimization of the antenna configuration to match the incident source distribution. This flexibility is essential for tailoring the optical system to the specific requirements of quantum communication protocols.
    \item \textbf{Practical Application:} The results are presented in a series of graphs that facilitate rapid and accurate evaluation of the antenna gain. This gain can then be substituted into the conventional range equation, making the model practical for real-world applications.
\end{enumerate}

From Klein and Degnan we have the antenna gain function:
\begin{equation}
    G(\alpha, \beta, \gamma, X, \lambda) = \left(\frac{4\pi A_\text{Tx}}{\lambda^2}\right)g(\alpha, \beta, \gamma, X)
\end{equation}
In this context, $\alpha$ and $\gamma$ are system parameters, the variable X represents off-axis distributions and $\beta$ incorporates both near-field and defocusing effects. The term $A_\text{Tx}$ represents the effective primary area. The factor $\left(\frac{4\pi A_\text{Tx}}{\lambda^2}\right)$ is a well-known upper limit on antenna gain, achieved when a circular aperture is uniformly illuminated and unobscured. So, $g$ represents the transmitter efficiency factor. In this work we will only focus on far-field cases and on-axis antennas such that the equation simplify:
\begin{equation}
    G_\text{Tx}(\alpha, \gamma, D_\text{Tx}, \lambda) = \left(\frac{\pi D_\text{Tx}}{\lambda}\right)^2 \frac{2}{\alpha^2} \left( \exp{\left(-\alpha^2\right)}-\exp{\left(-\alpha^2\gamma^2\right)} \right)^2
\end{equation}
In practice we choose $\gamma=0.2$ to be constant, which is a common choice in the design of antennas. Deriving the optimum aperture to beamwidth ratio for a general obscuration is a complex problem as it requires solving a transcendental equation. Using second order perturbation analysis Klein and Degnan obtained an approximate analytical solution (accurate within $\pm 1\%$ for $\gamma \leq 0.4$):
\begin{equation}
    \alpha \approx 1.12 - 1.30\gamma^2 + 2.12\gamma^4
\end{equation}
Then one can compute the directivity:
\begin{equation}
    D = \frac{G_\text{Tx}}{\left( \exp{\left(-2\alpha^2\gamma^2\right)}-\exp{\left(-2\alpha^2\right)} \right)}
\end{equation}
and deduce the half beamwidth angle:
\begin{equation}
    \theta_\text{HBW} = \sqrt{\frac{8}{D}}
\end{equation}
One can also compute the half beamwidth angle of the receiving telescope replacing $D_\text{Tx}$ by $D_\text{Rx}$. This is especially useful to compute the field of view of the receiving telescope before computing the amount of noise entering in the measurement system. So the field of view is:
\begin{equation}
\label{eq:omegafov}
    \Omega_\text{FOV} = \frac{\pi \theta_\text{HBW}^2}{2}
\end{equation}
Finally, knowing the standard error of the beam pointing error $\sigma$ enables us to compute the corrected gain of the antenna:
\begin{equation}
\label{eq:corr_tx_gain}
    G_\text{Tx-corrected} = \frac{8 \left( \exp{\left(-2\alpha^2\gamma^2\right)}-\exp{\left(-2\alpha^2\right)} \right)}{\theta_\text{HBW}^2 + 4\sigma^2}
\end{equation}
It is common to express the power ratio between the received field and the sent field with the transmitter gain obtained in Equation~\ref{eq:corr_tx_gain}, with $A_\text{Rx}$ the receiving aperture area and $r$ the distance between the two sources:
\begin{equation}
    \frac{P_\text{Rx}}{P_\text{Tx}} = \frac{G_\text{Tx-corrected} A_\text{Rx}}{4\pi r^2}
\end{equation}
As a consequence, we obtain the loss associated to diffraction and pointing error. All the other contributors computed in part \ref{app:error models} are independent and they are multiplied together with the following loss to compute the overall loss of the channel. \eqref{eq:loss_diff_bpe} can then be integrated in the quantum network performance optimization problem (Sec.~\ref{sec:Joint optimization: maximising network performance}): 
\begin{equation}
\label{eq:loss_diff_bpe}
    \eta_\text{diff-BPE} = \frac{G_\text{Tx-corrected} D_\text{Rx}^2}{16 d_\text{min}^2(L_0)}
\end{equation}

\section{Error models}
\label{app:error models}

\subsection{Channel Transmission Efficiency}
\label{appendix: channel transmission efficiency}
The channel transmission efficiency $\eta$ can be expressed as a product of component efficiencies:
\begin{equation}
    \eta(L_0) = \eta_{\text{diff-BPE}}(d_{\text{min}}(L_0)) \cdot \eta_{\text{atm}}(d_{\text{min}}(L_0)) \cdot \eta_{\text{cpl}}
\end{equation}
where $\eta_{\text{diff-BPE}}$ is the efficiency due to diffraction and beam-pointing errors, $\eta_{\text{atm}}$ is that due to atmospheric absorption and scattering and $\eta_{\text{cpl}}$ is due to inefficiencies in atmosphere-fiber coupling. The first two component efficiencies depend on the distance $L_0$ between adjacent ground nodes. $d_{\text{min}}$ is the minimum slant distance from the satellite to a ground station:
\begin{equation}
    d_{\text{min}}(L_0) = \sqrt{R_E^2 + \left(R_E+h\right)^2 - 2R_E \left(R_E+h\right) \cos\left(\frac{L_0}{2R_E}\right)}
    \label{eq:slant distance}
\end{equation}
where $h$ is the satellite altitude and $R_E$ is the Earth's radius. Note the formula is time independent, since here we only consider in the static symmetric case.

\subsection{Diffraction and pointing loss}
\label{appendix: Diffraction and pointing loss}
This loss is a fundamental consequence of the wave nature of light. As the photon beam propagates from the satellite over the distance $d_{\text{min}}$, it naturally spreads out. The diffraction loss quantifies the fraction of the beam's power that is missed by the finite-sized receiving telescope on the ground. In this work, instead of assuming a Gaussian beam profile we decide to model an Annular Gaussian beam which is obscured at the center and truncated on the wings by the telescope design (see Appendix \ref{appendix:Loss modelling: Optical Antenna Gain}). The loss due to pointing error is directly included in this model. This loss arises from the mechanical difficulty of perfectly tracking a fast-moving LEO satellite. Small pointing errors cause the beam to be slightly off-center from the receiver, reducing the collected power.
\begin{equation}
    \eta_{\text{diff-BPE}}(L_0) = \frac{G_\text{Tx-corrected} D_{\text{R}}^2}{16 d_{\text{min}}^2(L_0)}
    \label{eq:diffraction loss}
\end{equation}
where $D_R$ is the receiver telescope diameter, $G_\text{Tx-corrected}$ is the corrected transmitter gain \eqref{eq:corr_tx_gain} and $L_0$ is the distance between adjacent ground stations. According to \cite{lu2022micius} we decided to set a state of the art pointing error of the order of 2µrad up to 1µrad for the optimistic Scenario. In \cite{lu2022micius} they observe a pointing error of about 1.2 µrad in a QKD scheme, it is important to note that for entanglement distribution the APT (acquiring, pointing, and tracking) with two telescopes is harder to manage.

\subsection{Atmospheric absorption and scattering loss}
This term accounts for photons that are absorbed or scattered by molecules and aerosols in the Earth's atmosphere. The loss is greater when the satellite is at a lower elevation angle, as the photons must traverse a longer path through the atmosphere. This is modeled using the Beer-Lambert law:
\begin{equation}
    \eta_{\text{atm}}(L_0) = (\eta_{\text{zenith}})^{\sec\theta} \quad \text{with} \quad \cos\theta = \frac{h}{d_{\text{min}}} - \frac{d_{\text{min}}^2 - h^2}{2R_E d_{\text{min}}}
\end{equation}
where $\eta_{\text{zenith}}$ is the best-case transmittance when the satellite is directly overhead (at the zenith), it is wavelength dependent and $\sec\theta$ is the airmass factor that accounts for the increased path length at a zenith angle $\theta$.

\subsection{Coupling Loss}
\label{app:coupling loss}
This is a fixed efficiency factor that accounts for all other static losses in the system (internal losses of receiving and transmitting telescopes). This includes imperfections in the ground station's optical components, wavefront distortions due to atmospheric turbulence that are not corrected by adaptive optics, and the efficiency of coupling the collected light into the single-mode fibers required for the detectors. These factors mainly come from manufacturers and companies that build the components.

The budget for the total coupling loss is calculated from the following components:
\begin{enumerate}
    \item \textbf{Telescope and Optical Module Attenuation ($\eta_{\text{optics}}$):} A total budget of $-10\,\text{dB}$ is used for the receiver, transmitter, and optical module attenuation. This is $-1\,\text{dB}$ more than the budget in \cite{de2023satellite}, owing to the assumption of a $1\,\text{m}$ receiver telescope instead of $0.8\,\text{m}$.
    \item \textbf{Detector Efficiency ($\eta_{\text{det}}$):} We assume $80\%$ efficiency ($\approx -1\,\text{dB}$) based on the ID Quantique datasheet \cite{snspdidq}.
    \item \textbf{Fiber Coupling Efficiency ($\eta_{\text{fc}}$):} Based on modeling data (\cite{gruneisen2017modeling}, Figure4a), a conservative coupling efficiency of $25\%$ ($\approx -6\,\text{dB}$) is set.
\end{enumerate}
Thus, total coupling efficiency is $\eta_\text{cpl} = \eta_{\text{optics}}\times \eta_{\text{det}} \times \eta_{\text{fc}}$, corresponding to total coupling loss budget of $\sim 17\text{dB}$. This budget represents a state-of-the-art but conservative value, which is expected to be improved in future (We expect $\sim 10 \text{dB}$ for near future and $\sim 6\text{dB}$ in optimistic case). It should be noted that no system margin is included in this analysis, since we attempt to characterise a realistic achievable performance rather than a contractually guaranteed performance.

\subsection{Background Noise}
The background noise is quantified by $Y_0$, the probability of a detector registering a count from background light in a single pulse. This is derived from the mean number of background photons, $n_{\text{bar}}$, arriving per pulse.
\begin{equation}
n_{\text{bar}} = \frac{H_{\text{sky}} \cdot \Omega_\text{FOV} \cdot A_R \cdot (1-\gamma^2) \cdot \Delta\lambda \cdot \Delta T}{E_p}
\end{equation}
where $H_{\text{sky}}$ is the sky's spectral irradiance, $\Omega_{\text{FOV}}$ is the receiver's field of view, $A_R$ is the receiving area, $\Delta\lambda$ is the filter bandwidth, $\Delta T$ is the coincidence time window and $\gamma$ is the receiver terminal central obstruction factor (see Appendix \ref{appendix:e}).

The arrival of these photons follows Poisson statistics. The probability of detecting at least one background photon ($k \ge 1$) is $1 - P(k=0)$.
\begin{equation}
    Y_0 = 1 - P(k=0) = 1 - e^{-n_{\text{bar}}}
\end{equation}
For the satellite Scenario, $n_{\text{bar}}$ is extremely small, so we use the first-order Taylor expansion $e^{-x} \approx 1-x$, which gives the highly accurate approximation:
\begin{equation}
    Y_0 \approx n_{\text{bar}}
\end{equation}

We define $Y_{0A}$ and $ Y_{0B}$ to be the corresponding $Y_0$ for the each distinct path each one of the entangled photons takes from the satellite to the ground. For the time being, we assume symmetric channels and take $Y_{0A} = Y_{0B} = n_{\text{bar}}$.

\subsection{Mean photon number}
\label{app:mean photon number}
Photons are ideal for long-distance transmission due to their minimal environmental interaction, so their polarization is a particularly attractive degree of freedom due to its simplicity and compatibility with existing optical technologies. Spontaneous parametric down-conversion (SPDC) sources are thence highly suitable for applications like satellite-based quantum networks.
Assuming the use of a pulsed SPDC source, we know from \cite{Ma2012} that the probability to emit $n$ photon-pairs is:
\begin{equation}
    P(\mu,n) = \frac{(1+n)(\frac{\mu}{2})^n}{(1+\frac{\mu}{2})^{n+2}}
\end{equation}
with $\mu$ the average number of photons generated per pulse.
Also, let $Y_n$ represent the yield of an n-photon-pair, which is the conditional probability that a coincidence detection event occurs given that the source emits an n-photon-pair. $Y_n$ is primarily composed of two components: the background noise and the true signal. If we assume that the background counts are independent of the signal photon detection, then:
\begin{equation}
    Y_n = (1 - (1-Y_{0A})(1-\eta_A)^n)(1 - (1-Y_{0B})(1-\eta_B)^n)
\end{equation}
One can read $Y_n$ as the probability that Alice detects at least one event on her side and Bob detects at least one event on his side given the same pump pulse. The definition of the gain $Q_n$ of an n-photon-pair follows naturally to be the product of $P(\mu,n)$ and $Y_n$. Then, one can define the overall gain of the photon pairs (the probability of a coincident detection event given a pump pulse):
\begin{equation}
    Q_\mu = \sum_{n=0}^{\infty} P(\mu,n) Y_n
\end{equation}
\begin{equation}
    Q_\mu = 1 
    - \frac{1-Y_{0A}}{(1+\eta_A\frac{\mu}{2})^2} 
    - \frac{1-Y_{0B}}{(1+\eta_B\frac{\mu}{2})^2}
    + \frac{(1-Y_{0A})(1-Y_{0B})}{(1+\eta_A\frac{\mu}{2} + \eta_B\frac{\mu}{2} - \eta_A\eta_B\frac{\mu}{2})^2}
    \label{eq:qmu}
\end{equation}
Then one should evaluate the quality of the received entanglement. Making the hypothesis that the state can be represented as a Werner state \cite{Werner1989} as shown in \cite{caminati2006nonseparable}, one could derive the fidelity as a function of the QBER. From \cite{Ma2012} one gets the error rate as a function of $\mu$:
\begin{equation}
    \text{QBER} = e_0 - \frac{(e_0-e_d) \eta_A \eta_B \mu (1+\frac{\mu}{2})}
    {Q_\mu (1+\eta_A\frac{\mu}{2})(1+\eta_B\frac{\mu}{2})(1+\eta_A  \frac{\mu}{2}+\eta_B\frac{\mu}{2}-\eta_A \eta_B\frac{\mu}{2})}
\end{equation}
With $e_0$ the error rate of background counts (or any polarization uncorrelated coincidence) which is expected to be $50\%$. $e_d$ is the probability for a Bell state to hit the wrong detector, which characterizes the alignment and stability of the optical system. Then the start-fidelity $F_{init}$ can be written in terms of $\mu$ as:
\begin{equation}
    F_{init} = \frac{2-3\text{QBER}(\mu)}{2}
\end{equation}
It is known that in QKD we can find an optimal value of $\mu$ that maximizes the secure key rate. The optimization routine performs a similar role here.
In the two following equations, we propose a series expansion of $Q_\mu$ and $\text{QBER}$ in which we assume background counts to be small and we assume a high loss regime, which is reasonable in the satellite case. This is useful to improve reader intuition about the photon number trade off and for numerical verification. Indeed, it appears that increasing $\mu$ increase the coincidence rate but also the error rate so one have to optimize it. So, neglecting all the terms proportional to $\eta_A^x\eta_B^y, x+y>2)$, one get:
\begin{equation}
    Q_{\mu, \text{asy}} = \eta_A\eta_B \mu \left(1+\frac{3\mu}{2}\right)
\end{equation}
\begin{equation}
    QBER_\text{asy} = \frac{e_d + \frac{\mu}{2}(1+e_d )}{1+\frac{3\mu}{2}}
\end{equation}

 \section{Derivation of the Mean Background Photon Number}
 \label{appendix:e}
 This appendix provides a detailed, step-by-step calculation of the mean background photon number per pulse ($n_{\text{bar}}$) for the default simulation case, using the parameters defined in the summary table. The calculation begins with the general formula for the mean number of background photons collected by a telescope:%
 \begin{equation}
 n_{\text{bar}} = \frac{H_{\text{sky}} \cdot \Omega_\text{FOV} \cdot A_R \cdot (1-\gamma^2) \cdot \Delta\lambda \cdot \Delta T}{E_p}
 \end{equation}
 where $\Omega_\text{FOV}$ is the solid angle of the Field of View, $A_R$ is the area of the receiver telescope, and $E_p$ is the energy of a single photon. The spectral irradiance of sky is taken from \cite{er2005background} considering a full moon clear night scenario. We take the following default values from the simulation's parameter table, ensuring all are in base SI units:
 \begin{table}[htbp]
    \centering
    \begin{tabular}{|l|l|l|}
    \hline
        \textbf{Symbol} & \textbf{Description} & \textbf{Value (in SI Units)} \\
    \hline
        $H_{\text{sky}}$ & Spectral irradiance of sky background & $1.5 \times 10^{3} \text{ W m}^{-2} \text{sr}^{-1} \text{m}^{-1}$\\
        $\Omega_{\text{FOV}}$ & Full angle of the Field of View &  $8.87 \times 10^{-13}$ \text{ sr} \\
        $D_R$ & Diameter of the receiver telescope & 1.0 m \\
        $\Delta\lambda$ & Spectral filter bandwidth & $1 \times 10^{-9}$ m \\
        $\Delta T$ & Coincidence time window & $1 \times 10^{-9}$ s \\
        $\lambda$ & Wavelength of entangled photons & $7.8 \times 10^{-7}$ m \\
        $hc$ & Planck's Constant $\times$ Speed of Light & $1.9864 \times 10^{-25}$ J$\cdot$m \\
    \hline
    \end{tabular}
\end{table}

We first calculate the intermediate terms $\Omega_\text{FOV}$, $A_R$, and $E_p$. As detailed in Appendix \ref{appendix:Loss modelling: Optical Antenna Gain}, one can compute the gain of a telescope, the method remain the same for a receiving telescope. By computing the gain and the half beamwidth, one can deduce $\Omega_\text{FOV}$. For a small circular FOV, the solid angle is $\Omega = \frac{\pi \Delta\theta_D^2}{4}$. We keep the same hypothesis of $\gamma = 0.2$ so $\alpha \approx 1.07$ and the gain begins:
\begin{equation*}
    G_\text{Rx}(\alpha, \gamma, D_\text{Rx}, \lambda) = \left(\frac{\pi D_\text{Rx}}{\lambda}\right)^2 \frac{2}{\alpha^2} \left( \exp{\left(-\alpha^2\right)}-\exp{\left(-\alpha^2\gamma^2\right)} \right)^2 = \left(\frac{\pi}{7.8 \times 10^{-7}}\right)^2 \cdot 0.7088 \approx 1.15 \times 10^{13}
\end{equation*}
$\Delta\theta_D = \sqrt{2}\theta_\text{HBW}$ for SMF coupling, so the half angle and the solid angle of the field of view are given by:
\begin{equation*}
    \theta_\text{HBW} = \sqrt{\frac{8 \left( \exp{\left(-2\alpha^2\gamma^2\right)}-\exp{\left(-2\alpha^2\right)} \right)}{G_\text{Rx}}} = \sqrt{\frac{8 \cdot 0.811}{1.15 \times 10^{13}}} \approx 7.51 \times 10^{-7} \text{ rad}
\end{equation*}
\begin{equation*}
    \Omega_\text{FOV} = \frac{\pi (\sqrt{2}\theta_\text{HBW})^2}{4} \approx 8.87 \times 10^{-13} \text{ sr}
\end{equation*}

The area of the circular telescope aperture is $A_R = \pi \cdot (D_R/2)^2$.
\begin{equation*}
A_R = \pi \cdot \left(\frac{1.0 \text{m}}{2}\right)^2 = 0.25\pi \approx 0.7854 \text{ m}^2
\end{equation*}

The energy of a single photon is $E_p = hc/\lambda$.
\begin{equation*}
E_p = \frac{1.9864 \times 10^{-25} \text{ J}\cdot\text{m}}{7.8 \times 10^{-7} \text{ m}} \approx 2.547 \times 10^{-19} \text{ J}
\end{equation*}

Finally, we substitute all the calculated SI unit values into the main formula:
\begin{equation*}
n_{\text{bar}} = \frac{(1500 \text{ W m}^{-3} \text{sr}^{-1}) \cdot (8.87 \times 10^{-13} \text{ sr}) \cdot (0.7854 \text{ m}^2) \cdot (1-0.2^2) \cdot (1 \times 10^{-9} \text{ m}) \cdot (1 \times 10^{-9} \text{ s})}{2.547 \times 10^{-19} \text{ J}}
\end{equation*}
\begin{equation*}
n_{\text{bar}} = \frac{1.00 \times 10^{-27}}{2.547 \times 10^{-19}} \approx 3.94 \times 10^{-9}
\end{equation*}

\section{Maximum possible ground distance}
\label{appendix:f}
The goal is to derive the formula for the maximum possible ground distance, $L_0$, between two ground stations that can simultaneously maintain a line-of-sight to a single satellite. This maximum distance is constrained by a minimum practical elevation angle, $\phi_{\text{min}}$, below which communication is not feasible.

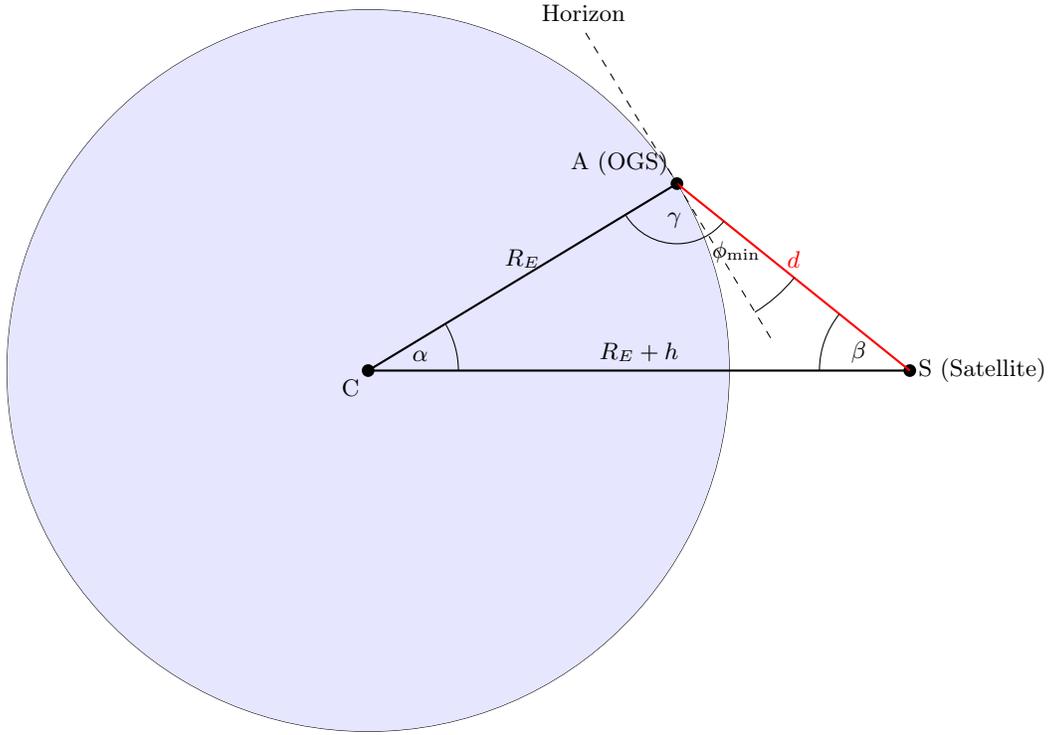
\begin{figure}[htbp]
    \centering
    \begin{tikzpicture}[
        scale=1.2,
        myR/.store in=\myR, myR=4, 
        myH/.store in=\myH, myH=2, 
        myPhiMin/.store in=\myPhiMin, myPhiMin=20 
    ]
        \pgfmathsetmacro{\myGamma}{90 + \myPhiMin}
        \pgfmathsetmacro{\mySinBeta}{((\myR) * sin(\myGamma)) / (\myR + \myH)}
        \pgfmathsetmacro{\myBeta}{asin(\mySinBeta)}
        \pgfmathsetmacro{\myAlpha}{180 - \myGamma - \myBeta}

        \coordinate (C) at (0,0);
        \coordinate (S) at (0:\myR+\myH);
        \coordinate (A) at (\myAlpha:\myR);
        \coordinate (HorizonPointAbove) at ($(A) + (\myAlpha+90:2)$); 
        \coordinate (HorizonPointBelow) at ($(A) + (\myAlpha-90:2)$); 

        \draw (C) circle (\myR);
        \fill[blue!10] (C) circle (\myR);
        \node at (C) [below left] {C};
        \fill (C) circle (2pt);

        \node at (S) [right] {S (Satellite)};
        \fill (S) circle (2pt);
        
        \node at (A) [above left] {A (OGS)};
        \fill (A) circle (2pt);

        \draw[thick] (C) -- (A) node[midway, above] {$R_E$};
        \draw[thick] (C) -- (S) node[midway, above] {$R_E + h$};
        \draw[thick, red] (A) -- (S) node[midway, above] {$d$};

        \draw[dashed] (HorizonPointBelow) -- (HorizonPointAbove) node[above, sloped] {Horizon};

        \pic [draw, angle radius=1.2cm, pic text=$\alpha$] {angle = S--C--A};
        \pic [draw, angle radius=1.2cm, pic text=$\beta$] {angle = A--S--C};
        \pic [draw, angle radius=0.8cm, pic text=$\gamma$] {angle = C--A--S};
        \pic [draw, angle radius=2cm, pic text=$\phi_{\text{min}}$] {angle = HorizonPointBelow--A--S};

    \end{tikzpicture}
    \caption{Geometric relationship for calculating maximum ground distance. $\gamma = 90^\circ + \phi_{\text{min}}$. The diagram is drawn to scale for the given angles.}
    \label{fig:geometry_lmax}
\end{figure}

We model the scenario using a 2D cross-section of the Earth. The geometry involves the Earth's center (C), a ground station (A), and the satellite (S) at its lowest visible point on the horizon, as defined by $\phi_{\text{min}}$. This is illustrated in Figure~\ref{fig:geometry_lmax}.

The derivation relies on solving the triangle CSA using the Law of Sines. The angle at the ground station, $\gamma$, is the sum of the 90-degree angle between the local horizon and the Earth's radius, and the minimum elevation angle $\phi_{\text{min}}$.
\begin{equation}
    \gamma = 90^\circ + \phi_{\text{min}}
\end{equation}
The sum of the angles in the triangle must be $180^\circ$ (or $\pi$ radians):
\begin{equation}
    \alpha + \beta + \gamma = 180^\circ \implies \alpha + \beta = 90^\circ - \phi_{\text{min}}
    \label{eq:angle_sum}
\end{equation}

The Law of Sines states that for any triangle, the ratio of the length of a side to the sine of its opposite angle is constant. For triangle CSA:
\begin{equation}
    \frac{\sin(\beta)}{R_E} = \frac{\sin(\gamma)}{R_E + h}
\end{equation}
We can solve for $\sin(\beta)$:
\begin{equation}
    \sin(\beta) = \frac{R_E \sin(\gamma)}{R_E + h}
\end{equation}
Substituting $\gamma = 90^\circ + \phi_{\text{min}}$ and using the identity $\sin(90^\circ + x) = \cos(x)$:
\begin{equation}
    \sin(\beta) = \frac{R_E \cos(\phi_{\text{min}})}{R_E + h}
    \label{eq:sin_beta}
\end{equation}

From the angle sum in Equation~\ref{eq:angle_sum}, we have $\alpha = 90^\circ - \phi_{\text{min}} - \beta$. To find $\alpha$, we first find $\beta$ by taking the arcsin of Equation~\ref{eq:sin_beta}:
\begin{equation}
    \beta = \arcsin\left(\frac{R_E \cos(\phi_{\text{min}})}{R_E + h}\right)
\end{equation}
Therefore, the central angle $\alpha$ is:
\begin{equation}
    \alpha = 90^\circ - \phi_{\text{min}} - \arcsin\left(\frac{R_E \cos(\phi_{\text{min}})}{R_E + h}\right)
\end{equation}
Using the trigonometric identity $\arccos(x) = 90^\circ - \arcsin(x)$, we can simplify the expression for $\alpha$:
\begin{equation}
    \alpha = \arccos\left(\frac{R_E \cos(\phi_{\text{min}})}{R_E + h}\right) - \phi_{\text{min}}
    \label{eq:alpha_final}
\end{equation}
The maximum ground distance $L$ is the total arc length covered by two ground stations, each at the edge of the satellite's visibility. This corresponds to an angle of $2\alpha$ at the Earth's center.
\begin{equation}
    L_0 = 2 \times (\text{arc length for } \alpha) = 2 R_E \alpha
\end{equation}
Substituting our final expression for $\alpha$ from Equation~\ref{eq:alpha_final}, we arrive at the desired formula:
\begin{equation}
    L_0 = 2 R_E \left[\arccos\left(\frac{R_E \cos(\phi_{\text{min}})}{R_E + h}\right) - \phi_{\text{min}}\right]
\end{equation}
Note that all angles ($\phi_{\text{min}}$ and the output of arccos) must be in radians for this final calculation.

\end{document}